\documentclass[final,3p]{elsarticle}

\usepackage{graphicx}
\usepackage{color}
\usepackage{epsfig}
\usepackage{booktabs}
\usepackage{latexsym}
\usepackage{amsmath}
\usepackage{eufrak}
\usepackage{hyperref}
\usepackage{subfigure}
\usepackage{ulem}
\usepackage{lettrine}
\usepackage{verbatim}
\usepackage{array}
\usepackage{varwidth}
\usepackage{mathptmx} 
\usepackage{setspace}
\biboptions{numbers,compress}
\journal{Computers and Fluids}

\begin{document}

\begin{frontmatter}

\title{A combined Lattice Boltzmann and Immersed Boundary approach for predicting the vascular transport of differently shaped particles}

\author[iit,cemec]{Alessandro Coclite}
\ead{alessandro.coclite@iit.it}

\author[dimeg,cemec]{Marco Donato de Tullio}
\ead{marcodonato.detullio@poliba.it}

\author[dimeg,cemec]{Giuseppe Pascazio\corref{cor}}
\ead{giuseppe.pascazio@poliba.it}

\author[iit]{Paolo Decuzzi\corref{cor}}
\ead{paolo.decuzzi@iit.it}

\cortext[cor]{Corresponding author}

\address[iit]{Laboratory of Nanotechnology for Precision Medicine, Fondazione Istituto Italiano di Tecnologia, Via Morego 30 -- 16163 Genova, Italy }

\address[cemec]{Centro di Eccellenza in Meccanica Computazionale (CEMeC), Politecnico di Bari, Via Re David 200 -- 70125 Bari, Italy}

\address[dimeg]{Dipartimento di Meccanica, Matematica e Management (DMMM), Politecnico di Bari, Via Re David 200 -- 70125 Bari, Italy}

\begin{abstract}

Modelling the vascular transport and adhesion of man-made particles is crucial for optimizing their efficacy in the detection and treatment of diseases. 
Here, a Lattice Boltzmann and Immersed Boundary methods are combined together for predicting the near wall dynamics of particles with different shapes in a laminar flow. For the lattice Boltzmann modelling, a Gauss-Hermite projection is used to derive the lattice equation; wall boundary conditions are imposed through the Zou-He framework; and a moving least squares algorithm accurately reconstructs the forcing term accounting for the immersed boundary. 
First, the computational code is validated against two well-known test cases: the sedimentation of circular and elliptical cylinders in a quiescent fluid. A very good agreement is observed between the present results and those available in the literature. Then, the transport of circular, elliptical, rectangular, square and triangular particles is analyzed in a Couette flow, at Re=20. All particles drifted laterally across the stream lines reaching an equilibrium position, independently of the initial conditions. For this large Reynolds number, the particle shape has no significant effect on the final equilibrium position but it does affect the absolute value and periodicity of the angular velocity. Specifically, elongated particles show longer oscillation periods and, most interestingly, larger variations in angular velocity. The longest particles exhibit a zero angular velocity for almost the whole rotational period. Collectively, this data demonstrates that the proposed approach can be efficiently used for predicting complex particle dynamics in biologically relevant flows. 
This computational strategy could have significant impact in the field of computational nanomedicine for optimizing the specific delivery of therapeutic and imaging agents.

\end{abstract}


\end{frontmatter}

\section*{Introduction}
\label{intro}

The intravenous administration of nanoconstructs for the precise delivery of therapeutic and imaging agents has been demonstrated to have potential in the fight against deadly diseases, such as cancer and cardiovascular diseases~\cite{peer2007,charalambos2010}. 
Nanoconstructs are man-made, biocompatible and biodegradable objects that transported by the blood flow can reach any location within the vascular network and release their therapeutic cargo thereof~\cite{adriani2012,deven2012,nabil2015}.
Over the last decade, nanoconstructs exhibiting different sizes, shapes, surface properties and, more recently, also mechanical stiffnesses have been presented~\cite{decuzzi2009,anselmo2015,palange2015,merkel2011,gratton2008}. 
The size can range from a few tens of nanometers to a few microns; the shape can be spherical, discoidal, cylindrical; the surface can exhibit a positive, negative or neutral electrostatic charge and can be decorated with a variety of moieties for specific cell recognition; and the stiffness can vary from that of cells to metals. Size, shape, surface and stiffness have been shown by the authors and others to significantly affect the vascular and extravascular behaviour of systemically injected nanoconstructs, and are therefore referred to as the \textit{4S design parameters}~\cite{palange2015}.
An incredibly large set of nanoconstructs would be identified by considering all possible combinations of these 4S parameters. It is therefore simply impractical to screen all of them for their biomedical properties by in vitro and in vivo testing. In this context, computational tools can help in reducing the screening time and cost as well as in identifying a sub-set of optimal nanoconstruct configurations to be, eventually, tested experimentally.

The authors have extensively employed computational tools for elucidating the mechanisms regulating the vascular transport and adhesion of nanoconstructs with different size, shape and surface properties~\cite{decuzzi2004,lee2009,lee2013,hossain2014,hossain2013}.
For instance, mathematical and computational analyses led to the in silico identification of sub-micron nanoconstructs with a discoidal shape as one of the best size/shape combinations for targeting the tumor vasculature. This was also demonstrated experimentally~\cite{adriani2012,deven2012,palange2015,decuzzi2010}. 
Indeed, multiple techniques have been proposed for modeling vascular flow and nanoconstruct transport in a variety of vascular districts, including direct numerical simulations (DNS)~\cite{joseph2002}, immersed finite element methods (IFEM)~\cite{lee2013,lee2014}, immersed boundary methods (IBM)~\cite{ICCFD2012}, and Isogeometric Analysis (IA)~\cite{hossain2014,hossain2013}. 
These approaches are all based on the discretization and numerical integration of the Navier-Stokes and continuity equations, and still remains challenging the modelling of vascular transport of multiple particles with complex shapes and mechanical properties. On the other hand, particle-based techniques, such as the Lattice Boltzmann method (LBM), provide simpler implementation, higher flexibility for a variety of applications and can be readily translated into parallel computing, thus allowing to handle complex flows and boundary conditions~\cite{ladd19941,ladd19942,succi2001}.
LBM uses the mesoscopic Boltzmann equation to determine macroscopic fluid dynamics and mass transport and has been already employed for modelling the vascular transport of particle suspensions and deformable cells~\cite{sun2003,krueger2011}.

In this work, the transport of particles with different shapes next to a rigid wall is considered. 
A combined LB-IB model is used for solving the fluid dynamics and then, estimating the forces over the immersed boundary.
This allows to solve the rigid-body transport problem for the immersed particle.
The rigid wall boundary conditions are imposed through the Zou-He framework to guarantee mass conservation and a Moving Least Squares (MLS)
algorithm is employed to accurately reconstruct the forcing term in the Boltzmann equation accounting for the presence of the 
boundary~\cite{vanella2009}.
For the LBM modelling, the lattice equation is derived on a rigorous mathematical basis by a Gauss-Hermite projection. 
The particle dynamics is simulated through a rigid-body-dynamics equations solver~\cite{ICCFD2012} weakly coupled with the flow solver. 

In the sequel, first the model is validated against two conventional test cases: the sedimentation of circular and elliptical cylinders in a quiescent fluid. Then, the transport of circular, elliptical, rectangular, square and triangular particles in a linear laminar flow is analyzed, at a fixed Reynolds number (Re=20). The flow conditions resemble the transport of particles in the vicinity of the vessel walls in the macro circulation.

\section{Method}
\label{sec:method}

\subsection{The lattice-Boltzmann method}
\label{LBM}
 
The evolution of the fluid is defined in terms of a set of $N$ discrete distribution functions $\{f_i\}$ ($i=0,\ldots,N-1$) which obey the dimensionless Boltzmann equation,
\begin{equation}
\label{evoleq}
{f_{i}({\bf x}+{\bf e}_{i}\Delta t, t+\Delta t)-f_{i}({\bf x}, t)=
-\frac{\Delta t}{\tau}[f_{i}({\bf x}, t)-f_{i}^{eq}({\bf x}, t)]}\, ,
\end{equation}
in which ${\bf x}$ and $t$ are the spatial coordinates and time, respectively,
$\{{\bf e}_{i}\}$ ($i=0,\ldots,N-1$) is the set of discrete
velocities, $\Delta t$ is the time step, $\tau$ is the relaxation
time given by the unique non-null eigenvalue of the collision term in the BGK-approximation~\cite{bgk}.
The kinematic viscosity of the flow is related to the single relaxation time $\tau$ as $\nu=(2\tau -1)\Delta\, t/c_s^2$ being $c_s=1/\sqrt{3}$ the reticular sound speed.
The moments of the distribution functions define the fluid density $\rho=\sum_i f_i$, velocity ${\bf u}=\sum_i f_i {\bf e}_i/\rho$ and the pressure $p=\rho\, c_s^2=c_s^2\sum_i f_i$.
The local equilibrium density functions (EDF) $\{f_i^{eq}\}$ ($i=0,\ldots,N-1$) are expressed by a Maxwell-Boltzmann (MB) distribution, as follows:
\begin{equation}\label{eqfunc}
f_i^{eq}({\bf x},t)=\omega_i\rho\left[1+\frac{1}{c_s^2}({\bf e}_i \cdot{\bf u})+
\frac{1}{2 c_s^4}({\bf e}_i \cdot{\bf u})^2-\frac{1}{2c_s^2}{\bf u}^2
\right]\, .
\end{equation}
On the two-dimensional square lattice with $N=9$ speeds ($D2Q9$)~\cite{d2q9}, the set of discrete velocities is given by:
\begin{equation}
{\bf e}_i= 
\begin{cases} 
(0,0)\, , &\mbox{if } i=0 \\ 
\Bigl(cos\bigl(\frac{(i-1)\pi}{2}\bigr)\, , sin\bigl(\frac{(i-1)\pi}{2}\bigr)\Bigr)\, , & \mbox{if } i=1-4\ \, , \\ 
\sqrt{2} \Bigl(cos\bigl(\frac{(2i-9)\pi}{4}\bigr)\, , sin\bigl(\frac{(2i-9)\pi}{4}\bigr)\Bigr)\, , & \mbox{if } i=5-8 \\ 
\end{cases} 
\end{equation}
with the weight, $\omega_i=1/9$ for $i=1-4$, $\omega_i=1/36$ for $i=5-8$,
and $\omega_0=4/9$.
Here we adopt a discretization in velocity space of the MB distribution based on the quadrature of a Hermite polynomial expansion of this distribution~\cite{shan06bis}. In this way it is possible to get a lattice equation that exactly recovers a finite number of leading order moments of the MB distribution.
In such a scheme it has been later realized that a regularization step \cite{latt06} is needed for $\tau \neq 1$ to re-project the post-collision distribution functions onto the Hermite space \cite{zhang06}. 
However, for practical purposes, such step is only  necessary when the Knudsen number Kn=Ma$/$Re, where Ma is the Mach number and Re the Reynolds, becomes larger than some value ($\approx 0.05$)~\cite{zhang06,colos10}. 
For completeness, Ma is defined as the ratio between a reference velocity of the flow and the sound speed, Ma$=u_{ref}/c_s$ and Re as the ratio between the inertia forces and the viscous ones, so that Re$=\frac{u_{ref}\, L_{ref}}{\nu}$, with $L_{ref}$ reference length, and $\nu$ the kinematic viscosity, respectively.    
In the present paper the value of Kn is always smaller than $10^{-3}$ so that the regularization step was not implemented. 

An effective forcing term accounting for the boundary presence, ${\cal F}_i$, can be included as an additional factor at the right hand side of equation~\eqref{evoleq},
\begin{equation}
\label{evoleqForced}
{f_{i}({\bf x}+{\bf e}_{i}\Delta t, t+\Delta t)-f_{i}({\bf x}, t)=
-\frac{\Delta t}{\tau}[f_{i}({\bf x}, t)-f_{i}^{eq}({\bf x}, t)]+\Delta t {\cal F}_i}\, .
\end{equation}
Following the argument from Guo et al.~\cite{guo2002}, also developed in~\cite{derosis20141,suzuki2015,wang2015}, ${\cal F}_i$ is given by:
\begin{equation}
\label{forcing}
{{\cal F}_i= \Bigl(1-\frac{1}{2\, \tau}\Bigr)\omega_i\Bigl[\frac{{\bf e_i}-{\bf u}}{c_s^2}+\frac{{\bf e_i}\cdot{\bf u}}{c_s^4}{\bf e_i}\Bigr]\cdot {\bf f_{lb}}}\, 
\end{equation}
where ${\bf f_{lb}}$ is the body force evaluated through the formulation by Favier et al.~\cite{pinelli2014} combined with the moving least squares reconstruction proposed by Vanella and Balaras~\cite{vanella2009}.
Due to the forcing term in equation~\eqref{evoleqForced}, the macroscopic quantities, given by the moments of the distribution functions, are obtained as:
\begin{eqnarray}
\label{rho}
\rho &=& \sum_i f_i\, , \\
\label{ustar}
\rho {\bf u} &=& \sum_i f_i {\bf e}_i+\frac{\Delta t}{2}{\cal F}_i\, .
\end{eqnarray}
It is proved that in such a framework one can recover the forced Navier--Stokes equation with second order accuracy~\cite{guo2002,pinelli2014}.
In the present model the forcing term accounts for the presence of an arbitrary shaped body into the flow-field, whereas the external boundaries of the computational domain are treated with the known-velocity bounce back conditions by Zou and He~\cite{zouhe1997}.

\subsection{Immersed boundary technique} 

In the Immersed-Boundary technique~\cite{peskin2002}, an obstacle in the flow can be considered as a collection of Lagrangian markers superimposed to the Eulerian withstanding fluid lattice. 
In order to account for the presence of the body, forcing terms are added to the governing equations. Here, the moving--least-squares (MLS) reconstruction of Vanella et al.~\cite{vanella2009} is employed to evaluate the forcing term, ${\bf f}_{lb}$.
%Those Lagrangian markers are in practice the centroid of the triangular elements which discretize the surface enclosing the immersed body. 
%The body force ${\bf f_{lb}}$ represents, in fact, the local force  exerted by the fluid on the surface of the immersed body. 
%Note that, in order to overcome the small pressure and density oscillations arising when using the linear interpolation of Fadlun et al.~\cite{fadlun2000} the moving--least-squares (MLS) reconstruction of Vanella et al.~\cite{vanella2009} is employed to evaluate ${\bf f}_{lb}$.
%
\begin{figure}
\centering
\includegraphics[scale=0.5]{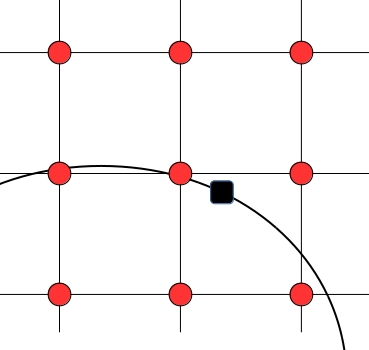}
\caption{Immersed-boundary technique schematic. The black square indicates the Lagrangian marker whereas the red circles represent the nine Eulerian points closest to the marker and involved in the forcing}
\label{IB}
\end{figure}
MLS reconstruction is the key ingredient to build a transfer function between the Eulerian lattice and the Lagrangian markers grid. 

Note that, among the different techniques proposed in the literature to enforce the boundary condition on a solid moving structure~\cite{chen2013,filippova1998,mei2000}, the MLS immersed boundary method should be preferred. In fact, even if interpolated bounce-back boundary condition are, in general, second order accurate, they are affected by several drawbacks. Precisely, dealing with an arbitrarily shaped geometry is a very hard task and then, due to the activation (or deactivation) of solid (or fluid) nodes, a refill procedure becomes necessary~\cite{derosis2014,li2016}. 
In turn the MLS-IBM gives a simple way to enforce boundary conditions keeping the second order accuracy given by the LBM-BGK scheme used.

Given a Lagrangian marker (with index $l$), nine Eulerian points are identified, namely the Eulerian points falling into the 
two-dimensional support domain, defined as a square with side equal to $r_w=2.6\, \Delta x$ and centred on the Lagrangian marker, see \textbf{Figure}~\ref{IB}. Given the solution at time level $n$, the velocity of the Lagrangian marker is evaluated as,
\begin{equation}
\label{uLag}
{{\bf U(x)}=\sum_{k=1}^9\, \phi^l_k({\bf x}){\bf u}_k}\, ,
\end{equation}
where $u_k$ indicates the velocity at the $k$-th Eulerian point associated with the marker and $\phi$ is the transfer operator 
obtained minimizing with respect to ${\bf a(x)}$ the following weighted L$_2$-norm:
\begin{equation}
\label{l2norm}
{J=\sum_{k=1}^9 W({\bf x}-{\bf x}^k}) [{\bf p}^T({\bf x}^k){\bf a(x)}-{\bf u}_k]^2\, .
\end{equation}
In the above equation, ${\bf p}^T$ is a basis function vector and ${\bf a(x)}$ a vector of coefficients such that 
$\sum_{k=1}^9 \, \phi^l_k({\bf x}){\bf u}_k\, ={\bf p}^T({\bf x}^k){\bf a(x)}$~\cite{vanella2009}, ${\bf x}$ 
the Lagrangian marker position, and $W({\bf x}-{\bf x}^k)$ a weight function.
In this work, a linear basis function, ${\bf p}^T=(1,x,y)$, is considered along with an exponential weight function:
\begin{equation}
\label{intKernel}
W({\bf x}-{\bf x}^k)= 
\begin{cases} 
e^{-(r_k/\alpha)^2}\, , &\mbox{if } r_k\leq 1 \\ 
0\, , & \mbox{if } r_k> 1\ \, , \\
\end{cases} 
\end{equation} 
where, $\alpha=0.3$ and $r_k$ is the distance between the Lagrangian point and the associated $k$-th Eulerian point normalised over the size of the support domain, $r_w$, $r_k=\frac{|{\bf x}-{\bf x}^k|}{r_w}$.
Now, for each Lagrangian marker the volume force ${\bf F}_l$ required to impose the boundary condition can be evaluated and then 
transferred to the Eulerian points into the support domain using the same functions used before.
% and properly scaled by a convenient factor $c_l$~\cite{vanella2009}.
Being ${\bf U}_{b,l}({\bf x})$ the desired velocity at the boundary related to the $l$-th triangle, the Lagrangian volume force is:
\begin{equation}
{{\bf F}_l {\bf (x)}=\frac{{\bf U}_{b,l} {\bf (x)}-{\bf U(x)}}{\Delta t}}\, ,
\end{equation}
and then the body force for the $k$-th Eulerian point is finally,
\begin{equation}
\label{bodyforce}
{{\bf f_{lb}}^k}=\sum_{l} c_l \phi^l_k {\bf F}_l \, .
\end{equation}
The scale coefficient $c_l$ is conveniently obtained by imposing the conservation of the total force acting on the fluid~\cite{vanella2009}.
The proposed algorithm concludes by inserting ${\bf f_{lb}}$ in equation~\eqref{forcing} and starting the next step by transporting the distribution functions ${\bf f}_i$ through equation~\eqref{evoleqForced}. In this way, the information given by the forcing term evaluated at the time level $n$ is transferred to the $n+1$-th time level. 
 
\subsection{Fluid-structure interaction} 
The total force and total moment acting on the immersed body are evaluated in time by integrating the viscous and pressure stresses over the body surface. 
Let $nl$ be the number of triangles composing the immersed surface and $l$ the triangular element index, one has:
\begin{eqnarray}
\label{forces}
{\bf F}(t)&=&\sum_{l=1}^{nl} (\overline{\tau}_l\cdot{\bf n}_l-p_l{\bf n}_l)\, S_l\ , \\
\label{momentum}
{\bf M}(t)&=&\sum_{l=1}^{nl} [{\bf r}_l \times (\overline{\tau}_l\cdot{\bf n}_l-p_l{\bf n}_l)]\, S_l\, ,
\end{eqnarray} 
where $\overline{\tau}_l$ and $p_l$ are the viscous stress tensor and the pressure evaluated in the centroid of the $l$-th triangle, ${\bf r}_l$ is the distance between the $l$-th Lagrangian marker and the centroid of the immersed body, ${\bf n}_l$ and $S_l$ are the outward normal unit vector and the area of the $l$-th triangle. The pressure and the velocity derivatives needed in equations~\eqref{forces} and~\eqref{momentum} are evaluated considering a probe in the normal positive direction of each triangle, the probe length being setted as $1.2\, \Delta x$, and using the MLS formulation described in the previous section.
Being $p^P_l$ the pressure at the probe, the pressure on the Lagrangian marker is evaluated considering the acceleration of the marker, $d_t {\bf u}_l$,
so that: $p_l=p^P_l+d_t {\bf u}_l\cdot{\bf n}_l$. The velocity derivatives evaluated at the probe are considered equal to the ones on the markers~\cite{yang2006}.
After the force and torque on the particle are determined, the translation and rotation of the particle are updated at each Newtonian dynamics time step, by an explicit second order Euler scheme. 
Given the computed values of ${\bf F}(t)$ and ${\bf M}(t)$, the linear and angular accelerations are obtained, ${\bf \dot{u}}(t)={\bf F}(t)/m$ and ${\bf \dot{\omega}}(t)={\bf M}(t)/I$, respectively, being $m$ the particle mass and $I$ the inertia moment. 
The linear and angular velocity are computed as:
\begin{eqnarray}
\label{ueuler}
{\bf u}(t)&=&\frac{2}{3}\, (2\, {\bf u}(t-\Delta t)-\frac{1}{2}\, {\bf u}(t-2\Delta t)+{\bf \dot{u}}(t)\ \Delta t)\, ,\\
\label{weuler}
{\bf \omega}(t)&=&\frac{2}{3}\, (2\, {\bf \omega}(t-\Delta t)-\frac{1}{2}\, {\bf \omega}(t-2\Delta t)+{\bf \dot{\omega}}(t)\ \Delta t)\, ,
\end{eqnarray}
with that $\Delta x=\Delta t=1$. A weak coupling approach between the fluid and the particle is implemented. It is known that this approach is not stable for large velocity variations~\cite{zhang2004}, but these cases are beyond the aim of the present work.
  
\section{Validation}
\label{Validation}

In this section two validation tests are presented to study the sedimentation of a circular and an elliptical particle in a narrow channel. 
The Froude, Fr, and Reynolds, Re, numbers are defined using the particle characteristic length, $L_{ref}$, and the sedimentation velocity 
of the particle in an infinite medium, $u_{Max}$. 
The definition of Fr is given in the following accordingly to the test case presented while Re is defined as, Re$=\frac{u_{Max}\, L_{ref}}{\nu}$, where $\nu$ the kinematic viscosity.

\subsection{Sedimentation of a circular particle}
\begin{figure}
\centering
\subfigure[$ $]{\includegraphics[scale=0.55]{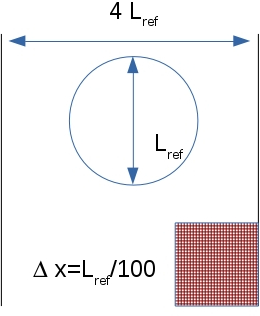}\label{Circ011}} \hspace{5.mm}
\subfigure[$ $]{\includegraphics[scale=0.20]{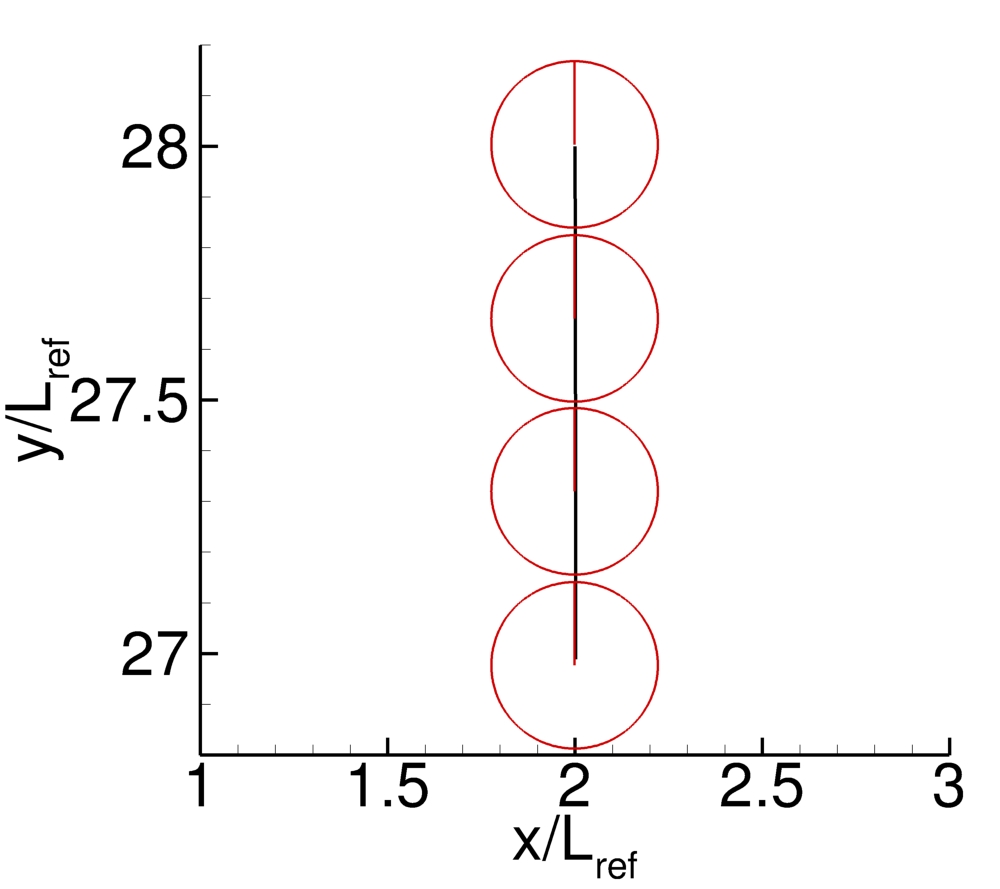}\label{Circ012}}\\
\subfigure[$ $]{\includegraphics[scale=0.47]{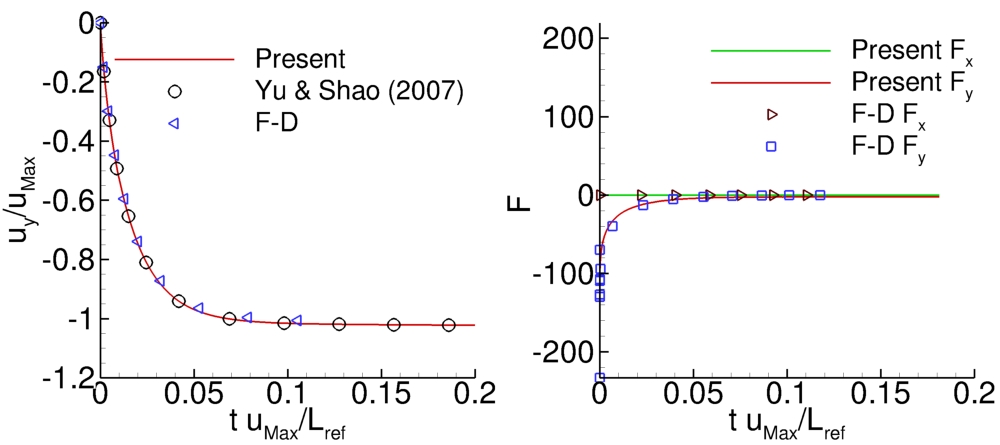}\label{Circ013}}
\caption{Sedimentation of a circular particle. (a) Schematic of the particle in a the narrow channel showing the computational grid. 
(b) Trajectory of the centroid of the circular particle (not in scale). 
(c) Centre of mass velocity (linear and angular) distributions over non-dimensional time (left plot), Force distributions (right plots). 
Lines are for the present LBM formulation results, circles correspond to the benchmark test by Yu\& Shao~\cite{yu2007}, triangles and squares are results obtained with a validated finite-difference approach~\cite{ICCFD2012,MDdTJCP2016}.}
\label{Circ01}
\end{figure}
We use the results from Yu and Shao~\cite{yu2007} about the sedimentation of a circular particle as benchmark solution. 
For comparison we use also the results obtained with a validated finite-difference method for the solution of the Navier-Stokes 
equation equipped with a MLS-IBM approach~\cite{ICCFD2012,MDdTJCP2016}. 
The computational domain is represented by a Cartesian uniform grid of size $4\, L_{ref}\times 30\, L_{ref}$, where the reference length is 
discretized with $100$ points and represents the diameter of the circle as shown in \textbf{Figure}~\ref{Circ011}. 
Initially the particle centroid is in $(x/L_{ref},y/L_{ref})=(2,28)$ with null velocity. The Reynolds number, defined through the terminal settling velocity, is Re$=0.1$ while the Froude number is Fr$=\frac{u_{Max}^2}{g\, L_{ref}}=1398.3$, $g$ corresponding to the gravitational acceleration. 
The settling velocity, $u_{Max}$, is such that $\Delta\, t=u_{Max}/L_{ref}=10^{-6}$ and $\tau$ is kept equal to unity in order to have better stability and accuracy.
The ratio between the solid and the fluid densities is $\rho_s/\rho_l=1.2$. 
The cylinder surface is uniformly discretized so that the ratio between the solid and the fluid meshes is about $0.3$.
\textbf{Figure}~\ref{Circ012} shows the trajectory of the circular particle: four representative positions of the particle are depicted (red circles) indicating a very small angular oscillation around the equilibrium position ($\theta=\theta_{0}=0$).
The comparison with data published by Yu\&Shao~\cite{yu2007} and those obtained by de Tullio et al.~\cite{ICCFD2012} is shown 
in~\textbf{Figure}~\ref{Circ013} where the left panel gives the vertical velocity components and the right panel gives the two components of the total force acting on the particle, ${\bf F}\,\rho\,{\bf u}_{Max}^2\, L_{ref}$.
An estimation of the accuracy between the present approach and the finite difference one is given considering the 
point in which the largest difference is observed in the sedimentation velocity and evaluating the relative error $\epsilon=(u_{LBM}-u_{F-D})/u_{F-D}$. The largest displacement between the two solution is given for $t\times u_{Max}/L_{ref}=0.03$ and results that $\epsilon=5.04\times 10^{-3}$.
\begin{figure}
\centering
\includegraphics[scale=0.3]{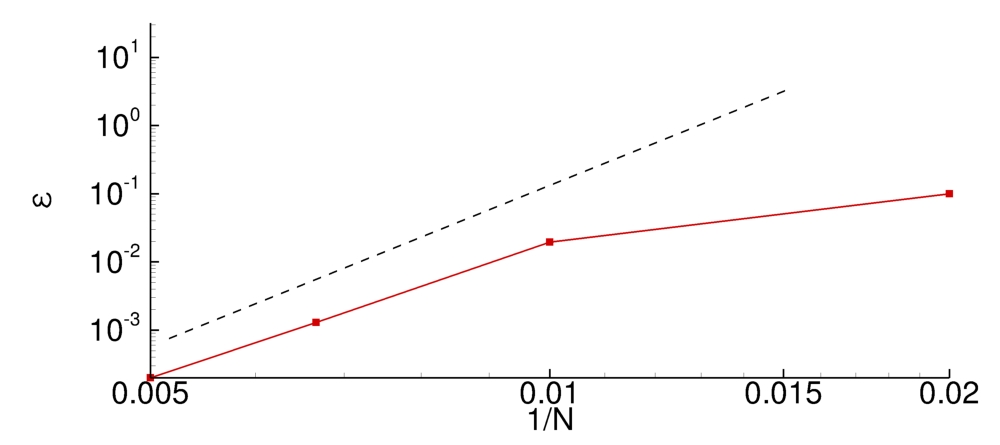}
\caption{Mesh-refinement study on the sedimentation terminal velocity; the dashed line corresponds to the second order convergence rate.}
\label{convergence}
\end{figure}
This test case has also been used to perform a mesh-refinement convergence analysis of the computational method.
The sedimentation velocity of the circular particle is computed as a function of the mesh discretization.
In particular, the particle diameter is discretized with $N=$50, 100, 150, and 200 points; while the ratio between the solid and the fluid meshes is kept constant and equal to about 0.3 $\Delta x$; also the relaxation parameter, $\tau$, is kept equal to unity for all simulations. The relative error for the sedimentation velocity $\epsilon=(u_{N} - u_{e})/u_{e}$ is evaluated with respect to the velocity obtained for $N=400$ points and $\tau=1$.
\textbf{Figure}~\ref{convergence} gives the behaviour of $\epsilon$ versus the mesh size $1/N$ showing that the error is already smaller than 10$\%$ for N=50 and the method is
second-order accurate.

\subsection{Sedimentation of an elliptical particle}
\begin{figure}
\centering
\subfigure[$ $]{\includegraphics[scale=0.45]{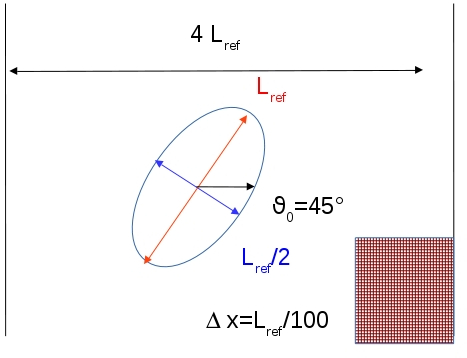}\label{Ell121}}
\subfigure[$ $]{\includegraphics[scale=0.200]{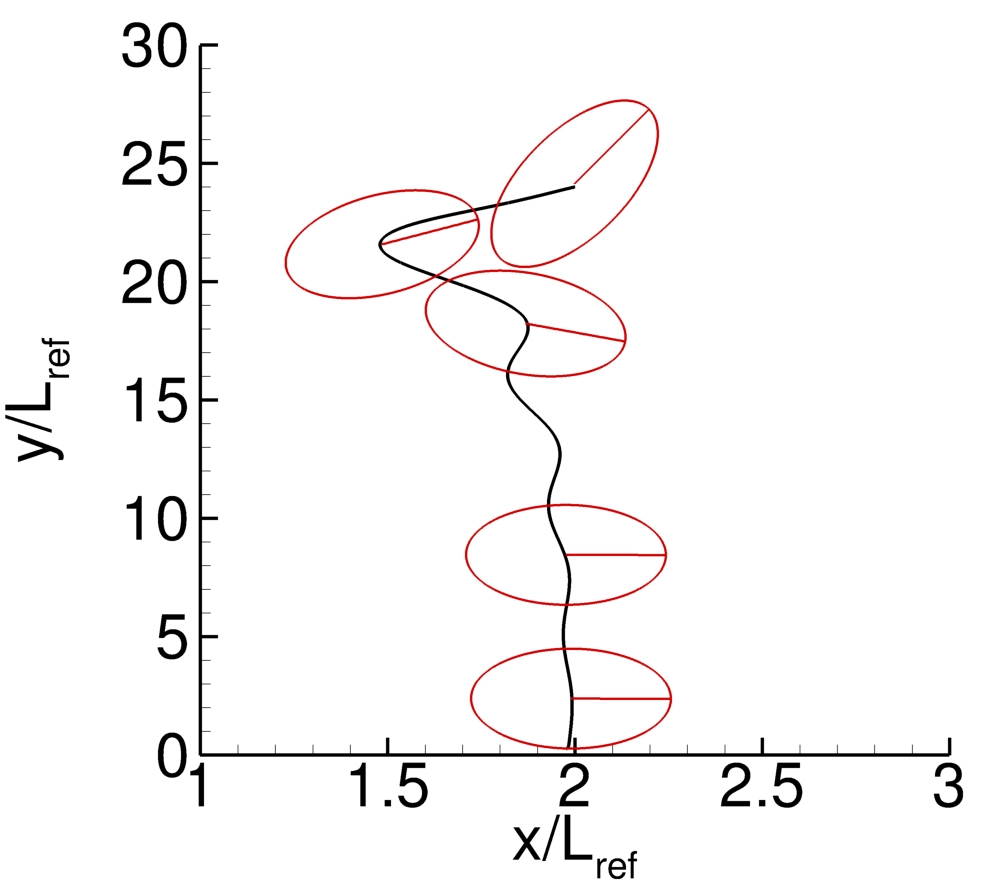}\label{Ell122}}\\
\subfigure[$ $]{\includegraphics[scale=0.47]{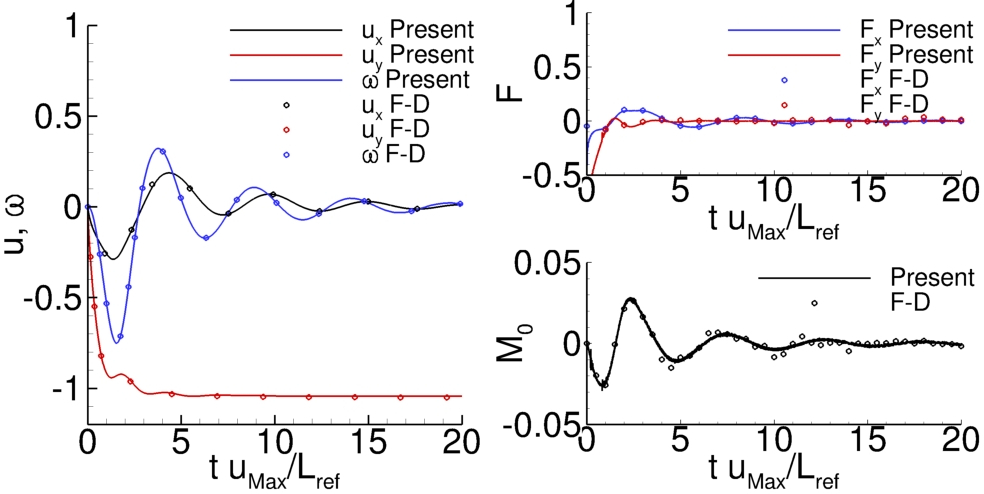}\label{Ell123}}
\caption{Sedimentation of an elliptical particle. (a), Schematic of the particle in a the narrow channel showing the computational grid. (b), Trajectory of the centroid of the elliptical particle (not in scale). (c), Centre of mass velocity (linear and angular) distributions over non-dimensional time (left plot), Force and moment distributions (right plots). Lines are for the present LBM formulation results, symbols are related to a validated finite-difference approach~\cite{ICCFD2012}.}
\label{Ell12}
\end{figure}
Let us consider a second validation test, namely, the sedimentation of an elliptical particle, for which we use the results by 
Xia et al.~\cite{xia2009} as benchmark solution. 
The computational domain is represented by a Cartesian uniform grid of $30\, L_{ref}\times 4\, L_{ref}$ where the reference length is the major axis of the ellipse, as shown in \textbf{Figure}~\ref{Ell121}, and is discretized with $100$ points.
The initial linear and angular velocity are null while the initial position of the centroid is $(x/L_{ref},y/L_{ref})=(2,24)$ and the particle starts falling with initial angle $\theta_0=\pi/4$. The Reynolds number is Re$=12.5$ while the Froude number is Fr$=\frac{u_{Max}}{\sqrt{L_{ref}\ g}}=0.126$. The ratio between the solid and fluid densities is $\rho_s/\rho_l=1.1$.  
The settling velocity, $u_{Max}$, is such that $\Delta\, t=u_{Max}/L_{ref}=10^{-4}$ and $\tau$ is again kept equal to unity in order to have the better stability and accuracy.
The surface of the elliptical particle is uniformly discretized so that the distance between two Lagrangian markers is 
about $0.3\, \Delta x$.  
\textbf{Figure}~\ref{Ell122} provides the trajectory of the centre of mass in the x-y plane: the particle in
five representative time instants is drown. After few initial fluctuations the particle finds its equilibrium position and settles in the middle of the channel with an horizontal inclination.
\textbf{Figure}~\ref{Ell123} shows the comparison with the finite difference Navier-Stokes IB method~\cite{ICCFD2012,MDdTJCP2016}. 
The maximum displacement of the sedimentation velocity with respect to the solution of such a method is found for 
$t\times u_{Max}/L_{ref}=4.7$ and the corresponding relative error is equal to $\epsilon=3.07\times 10^{-4}$.
The comparison of the present solution with that of de Tullio et al~\cite{ICCFD2012,MDdTJCP2016} is provided 
in \textbf{Figure}~\ref{Ell123} also for the two components of the force and the moment. 
While the relative error found for the linear and angular velocitis is represented by the value of $\epsilon$ given above, 
for the moment, see \textbf{Figure}~\ref{Ell123}~(right-bottom plot) a larger value is found, namely, $1.03\times 10^{-3}$, 
at the same non-dimensional instant. 
From this comparison  we conclude that the proposed model is validated versus a different fluid solver with the same IB-FSI procedure.

\begin{figure}
\centering
\includegraphics[scale=0.4]{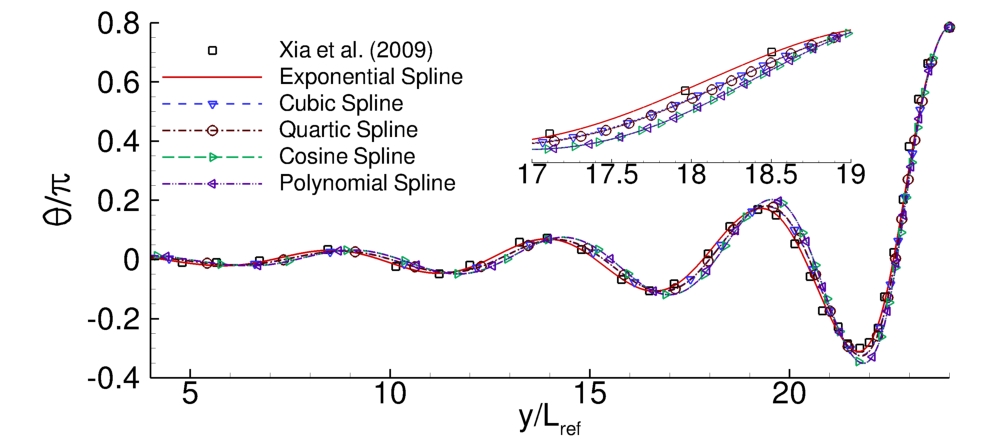}
\caption{Comparison between the five interpolating kernels and the benchmark data provided by Xia et al.~\cite{xia2009}.}
\label{EllSpline}
\end{figure}

To further validate the model a comparison among several interpolating kernels was carried out.
\textbf{Figure}~\ref{EllSpline} shows a direct comparison among the five different interpolating functions: 
\begin{itemize}
\item exponential kernel, used here (see \textbf{Eq.}~\eqref{intKernel}, 
\item cubic kernel:
\begin{equation}
\label{cubic}
W_{Cubic}(r_k) =
\begin{cases} 
\frac{2}{3}-4\, r_k^2+4\, r_k^3\, , &\mbox{if } r_k\leq 1 \\ 
0\, , & \mbox{if } r_k> 1\ \, \\
\end{cases}\, ,\\ 
\end{equation}
\item quartic kernel:
\begin{equation}
\label{quartic}
W_{Quartic}(r_k) = 
\begin{cases} 
1-6\, r_k^2+8\, r_k^3-3\, r_k^4\, , &\mbox{if } r_k\leq 1 \\ 
0\, , & \mbox{if } r_k> 1\ \, \\
\end{cases}\, ,\\
\end{equation}
\item cosine kernel:
\begin{equation}
\label{cosine}
W_{Cosine}(r_k) = 
\begin{cases} 
\frac{1}{4}[1+cos(\frac{\pi}{2}r_k)]\, , &\mbox{if } r_k\leq 1 \\ 
0\, , & \mbox{if } r_k> 1\ \,  \\
\end{cases}\, ,\\
\end{equation}
\item polynomial kernel:
\begin{equation}
\label{polynomial}
W_{Polynomial}(r_k) = 
\begin{cases} 
\frac{1}{8}(3-2\, r_k+\sqrt{1+4\, r_k-4\, r_k^2})\, , &\mbox{if } r_k\leq 0.5 \\ 
\frac{1}{8}(5-2\, r_k-\sqrt{-7+12\, r_k-4\, r_k^2})\, , &\mbox{if } \ 0.5 < r_k\leq 1 \\ 
0\, , & \mbox{if } r_k> 1\ \, 
\end{cases}\, ,\\
\end{equation} 
\end{itemize}
and the benchmark data provided by Xia et al.~\cite{xia2009}.
For all different functions, the orientation $\theta/\Pi$ of the particle major axis is compared during the sedimentation process. As documented in \textbf{Figure}~\ref{EllSpline}, the difference among the five interpolating kernels and the benchmark is minimal.
The maximum difference between the exponential function and the data by Xia et al.~\cite{xia2009} occurs at $y/L_{ref}=18$ with a relative error 
of $\epsilon=1.5\times 10^{-3}$.
For the other interpolating kernels larger differences are observed, specifically, $\epsilon_{Cubic}=1.7\times 10^{-2}$, $\epsilon_{Quartic}=1.7\times 10^{-2}$, $\epsilon_{Cosine}=2\times 10^{-2}$, and, $\epsilon_{Polynomial}=2\times 10^{-2}$.
The exponential function of \textbf{Equation}~\eqref{intKernel} is used in the sequel.

\section{Transport of differently shaped particles in shear flow}
\label{transportRes}
\subsection{Description of the numerical experiment}
\begin{figure}
\centering
\subfigure[]{\includegraphics[scale=0.4]{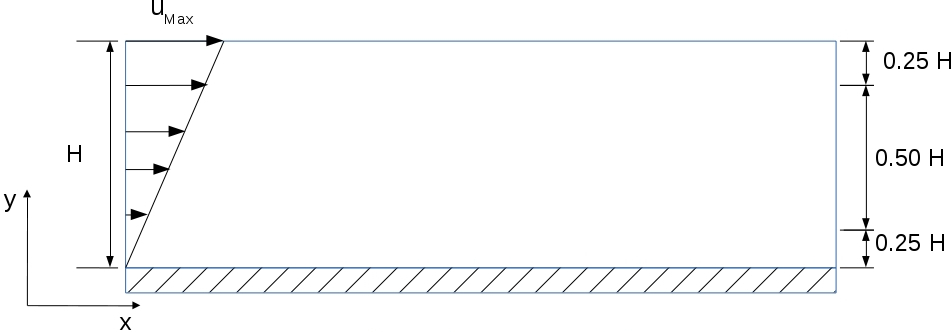}\label{transpSchematic0}}\\*[5mm]
\subfigure[]{\includegraphics[scale=0.5]{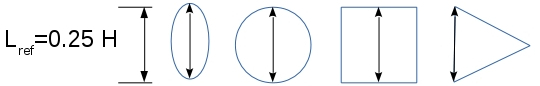}\label{transpSchematic1}}
\caption{Particle transport in a linear laminar flow. (a) Schematic representation of the computational domain. (b) Particle geometries and definition of $L_{ref}$}.
\label{transpSchematic}
\end{figure}
Here, two infinite planes in relative motion with respect to each other are considered: at $y/H=0$ the plane is fixed while and $y/H=1$ the plane is moving with a constant velocity along the $x$ direction 
at $u_{Wall}=u_{Max}$, as shown in \textbf{Figure}~\ref{transpSchematic0}.
No-slip boundary conditions are imposed on both planes.
The computational domain is $[0,10\, L_{ref}]\times[0, H]$, 
the height of the domain being $H=4\ L_{ref}$. 
The reference length, $L_{ref}$, definied in \textbf{Figure}~\ref{transpSchematic1}, is discretized with $100$ points. 
The flow-field is initialized with the planar Couette solution: $u(x)=u_{Max}\, y/H$.\\
Simulations have been performed considering a neutrally buoyant particle placed initially ($t=0$) at either $y/H_{t=0}=0.25$ or 0.5 and 0.75, with a null-velocity.
In the present case, four different particle shapes are considered, namely circular, elliptical, square, and triangular (\textbf{Figure}~\ref{transpSchematic1}). 
The surface of the particles is uniformly discretized so that the ratio between the solid and the fluid meshes is about
$0.3$. 
The Reynolds number, defined as $u_{Wall} H / \nu$, is fixed to Re$=20$, while the particle Reynolds number~\cite{feng19942} is given by Re$_p=$Re$\, L_{ref}/H$, equals $5$. $u_{Max}$ is chosen in order to have $\Delta\, t=10^{-4}$.  
Note that, once fixed $L_{ref}$, the particle areas differ one from each other.
With respect to the circular particle the normalized areas are: $0.5$, $1.27$, and $0.55$ respectively for the elliptical, square, and triangular particle. It should be here emphasized that given that small Knudsen number (Kn$\ll$1) and large Reynolds number (Re=20) considered in this work thermal fluctuations effects are neglected. 
In order to observe the interaction between the continue hydrodynamic forces and the momentum transferred by Brownian scattering the particle characteristic size must be lower than $1\, nm$. For this reason, thermal diffusion becomes relevant for Reynolds numbers around $10^{-3}$ and Knudsen numbers around $1$~\cite{fullstone2015,tan2012}.

\subsection{Transport mechanism for the circular particle in a linear laminar flow}
\begin{figure}
\centering
\subfigure[ ]{\includegraphics[scale=0.15]{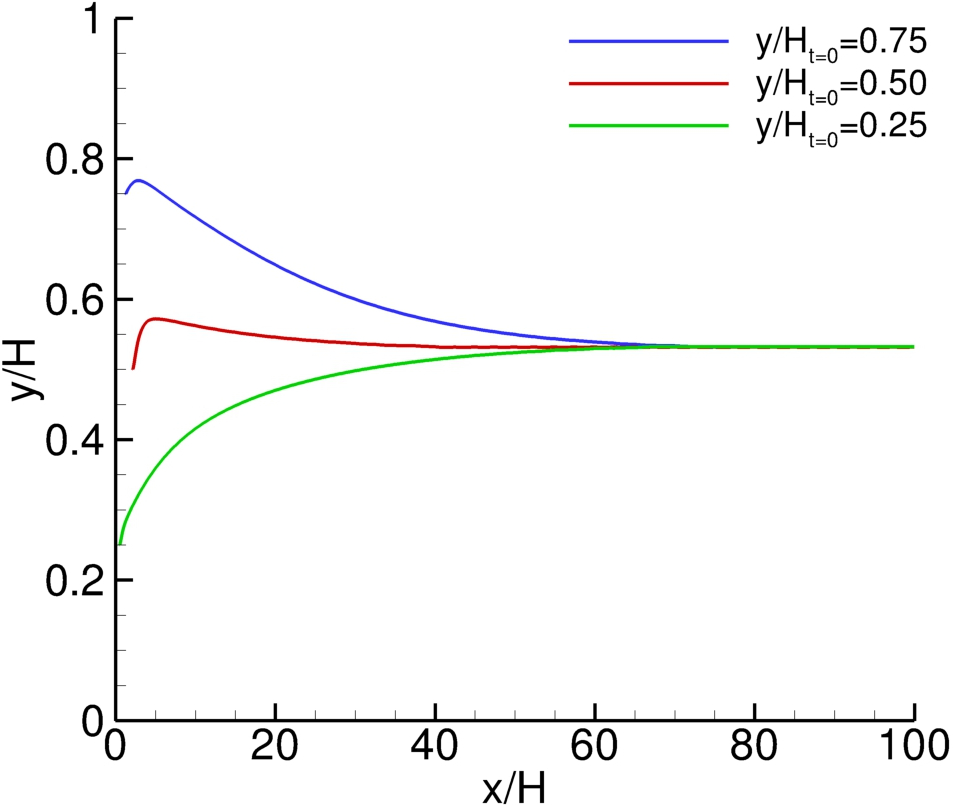}\label{CircleTraj}}\\
\subfigure[$y/H_{t=0}=0.25$]{\includegraphics[scale=0.15]{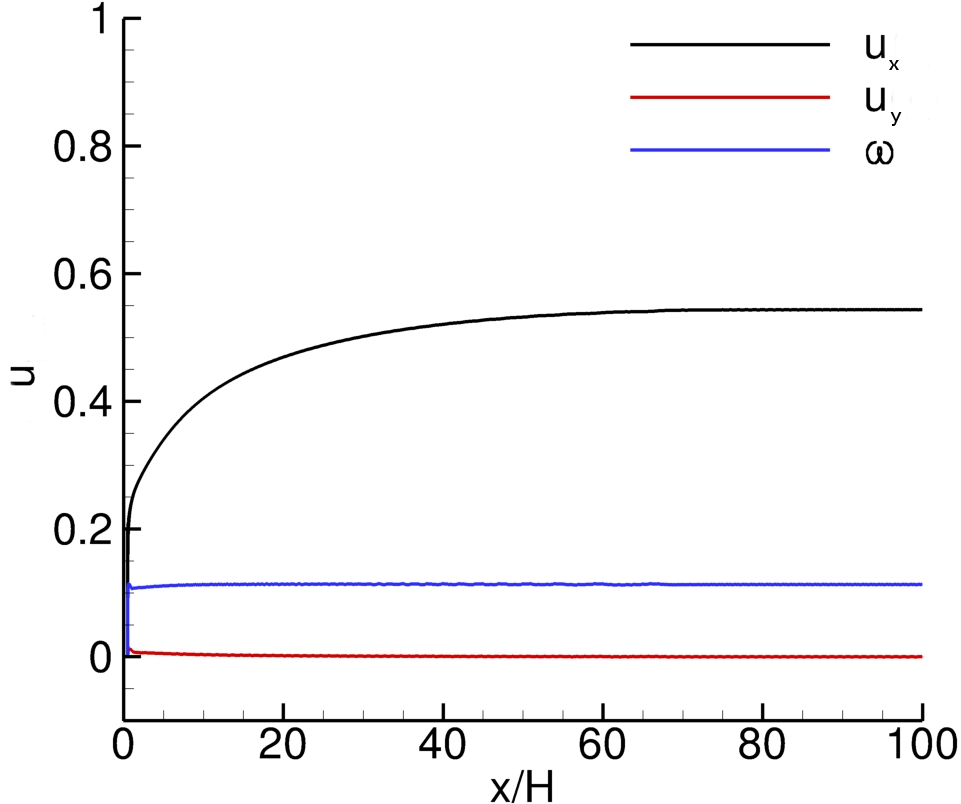}
\label{CircleVel025}}
\subfigure[$y/H_{t=0}=0.50$]{\includegraphics[scale=0.15]{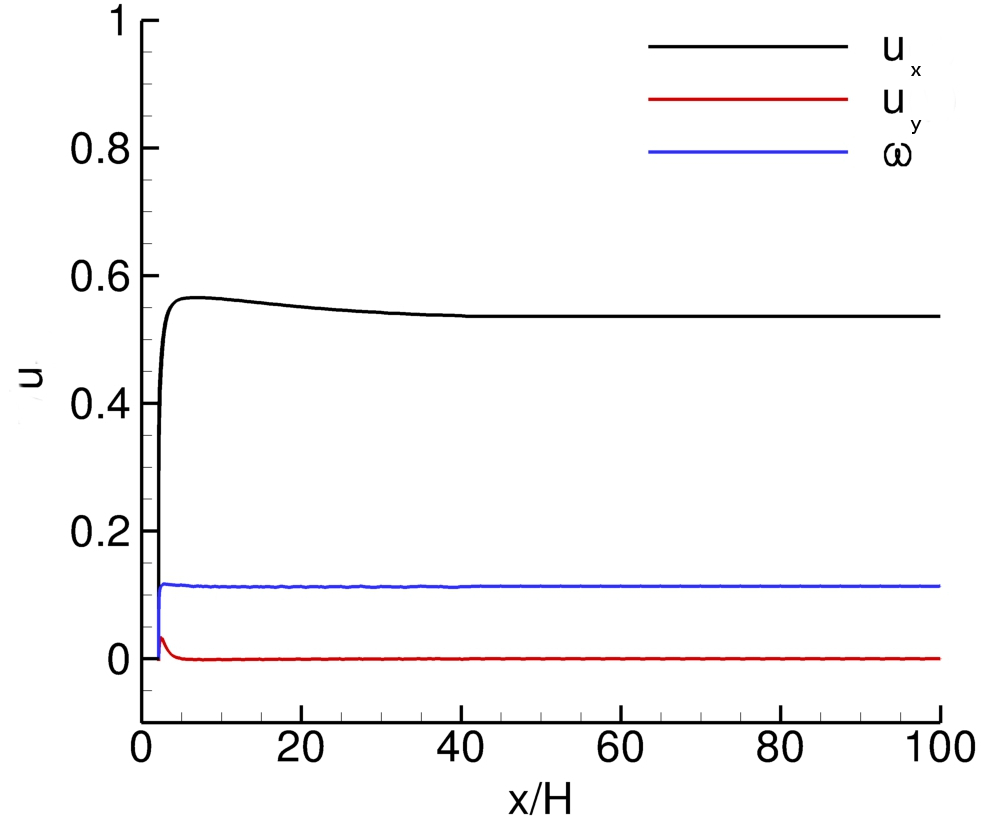}
\label{CircleVel050}}
\subfigure[$y/H_{t=0}=0.75$]{\includegraphics[scale=0.15]{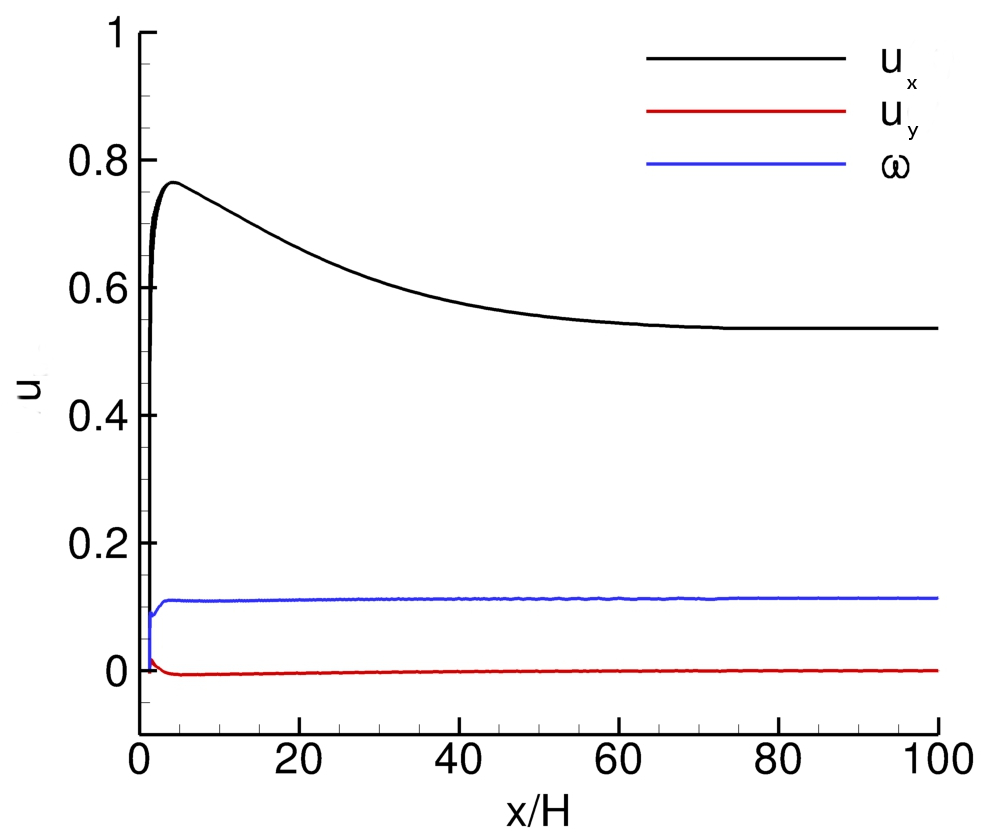}
\label{CircleVel075}}\\
\subfigure[ ]{\includegraphics[scale=0.15]{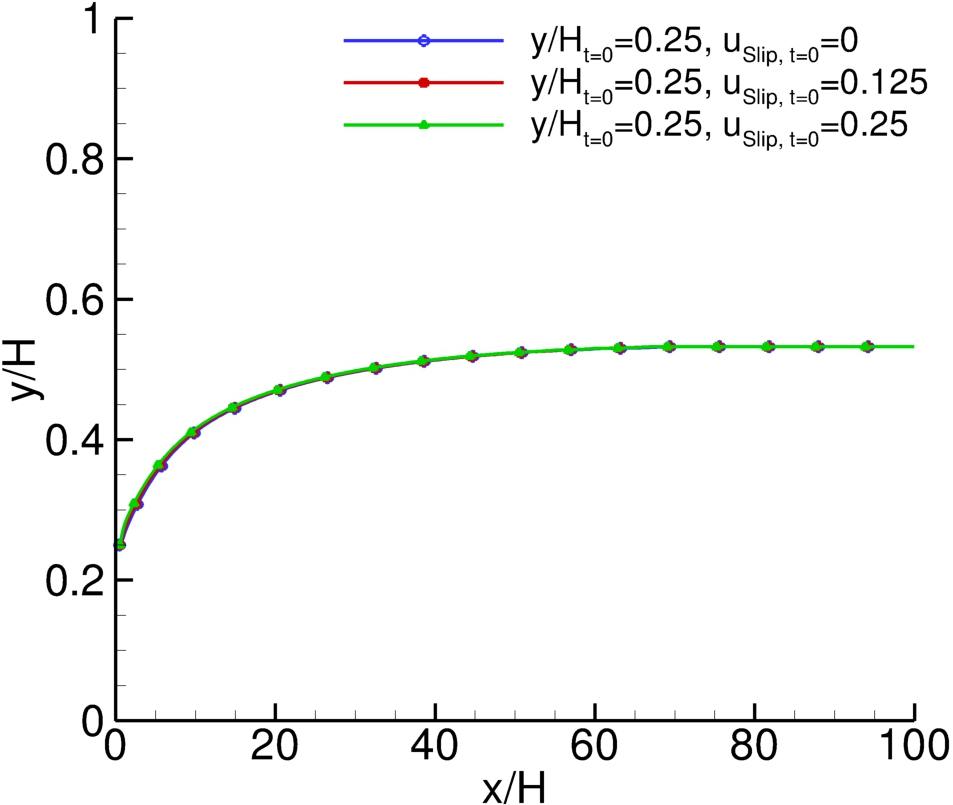}\label{TrajCircVel}}\\
\subfigure[ ]{\includegraphics[scale=0.15]{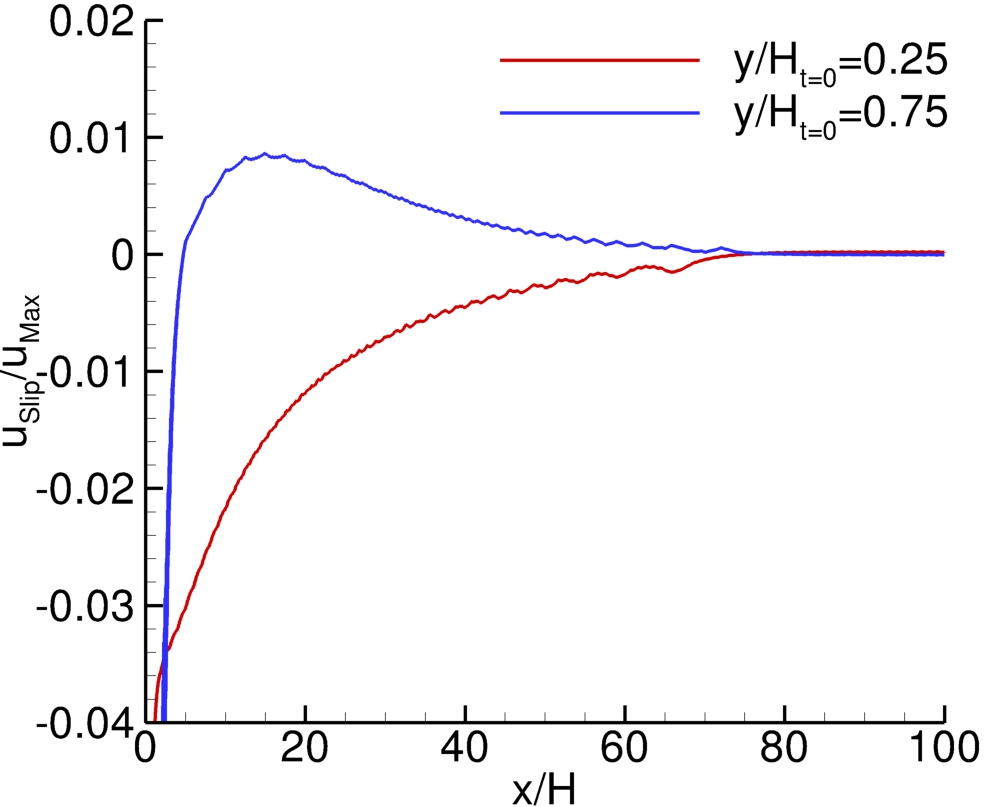}\label{uSlipU}}
\subfigure[ ]{\includegraphics[scale=0.15]{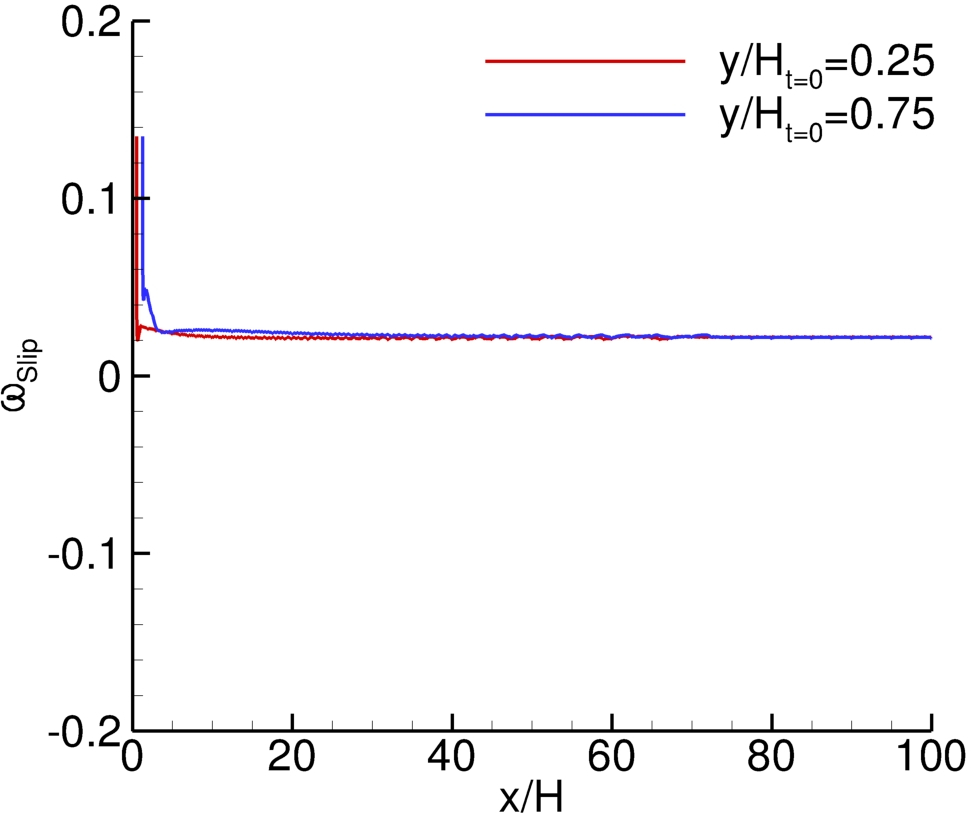}\label{uSlipOmega}}
\caption{Circular particle lateral migration. 
(a): Trajectory of the circular particle when released in $y/H_{t=0}=0.25,\ 0.50,\ 0.75$. (b), (c), (d): Linear and angular velocity distributions evaluated in the three different releasing position. (e) Trajectory obtained with different initial velocity. (f) Linear slip velocity distributions. (g) Angular slip velocity distributions.}
\label{CircleAll}
\end{figure}
\textbf{Figure}~\ref{CircleAll} shows the trajectories of the centroid of the circular particle initially located 
at different heights within the flow domain, namely $y/H_{t=0}=0.25,\ 0.50,\ 0.75$. 
As the particle moves along $x$ pushed by the flow, it also tends to drift laterally across the streamlines seeking for a stable equilibrium position. 
The particle released at $y/H_{t=0}=0.25$ moves towards higher velocities zones whereas the particle released at $y/H_{t=0}=0.75$ moves towards flow regions at lower speeds, as compared to the initial locations.  
The particle released at $y/H_{t=0}=0.50$ exhibits a minor jump towards the wall which is rapidly followed by a steady drift up to its final equilibrium position.
This is documented in \textbf{Figure}~\ref{CircleTraj} for all the three cases. Notably, the final stable equilibrium position is independent of the initial locations and coincides with $y/H=0.5$. 
It seems, in fact, that the centre of the flow domain represents an attractor for the particle, in agreement with previous data obtained via numerical simulations~\cite{feng19942,saffman1965}.
The particle velocities along $x$ are plotted for the three different cases in \textbf{Figures}~\ref{CircleVel025},~\ref{CircleVel050},~\ref{CircleVel075}. 
Note that $u_x$ grows from zero rapidly tending to $0.539\, u_{Max}$ which is about the undisturbed flow velocity at 
the equilibrium position.
The vertical velocity component is about three orders of magnitude lower than $u_x$, demonstrating that lateral drifting is a 
slower process. 
At the equilibrium, the particle has a finite angular velocity, $\omega_{eq}\times L_{ref}/u_{Max}=0.117$ 
which is comparable to one-half of the constant shear rate of the undisturbed flow, 
$\displaystyle{\frac{1}{2}} \, \dot\gamma = \displaystyle{\frac{1}{2} \frac{d u_x}{d y}} =  0.125$,
in agreement with results of Feng et al.~\cite{feng19942}. 
Moreover, it is important to observe that the equilibrium position is not affected by the initial velocity.
This is observed in \textbf{Figure}~\ref{TrajCircVel} showing the trajectories of the centroid for a particle released 
at $y/H=0.25$ with $u_{t=0}=0,\ 0.125,\ 0.25$: trajectories are perfectly superimposed one over each other and have again the center of the flow domain as the equilibrium position~\cite{feng19942,saffman1965}. 

\begin{figure}
\centering
\subfigure[t=10]{\includegraphics[scale=0.18]{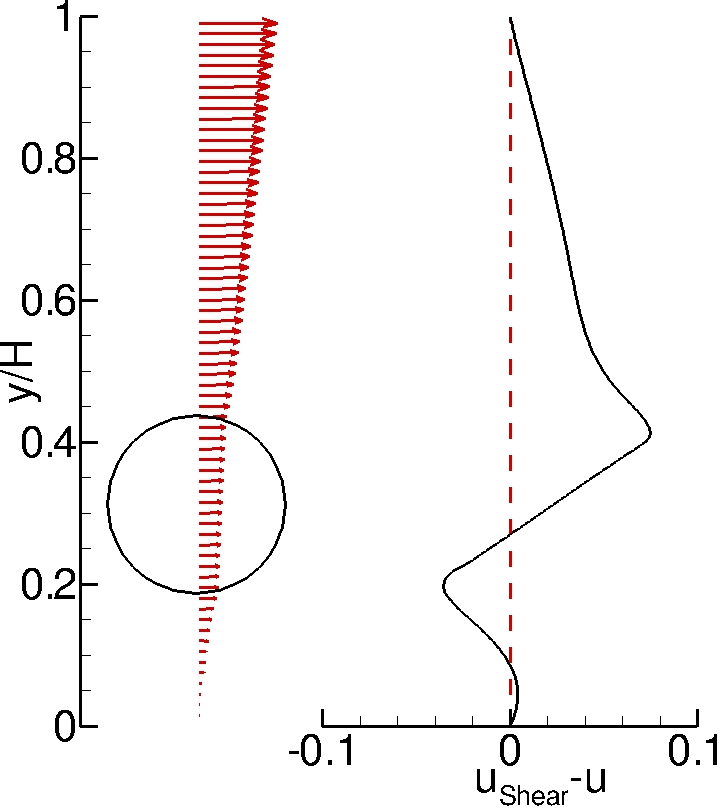}\label{profVel10}}
\subfigure[t=50]{\includegraphics[scale=0.18]{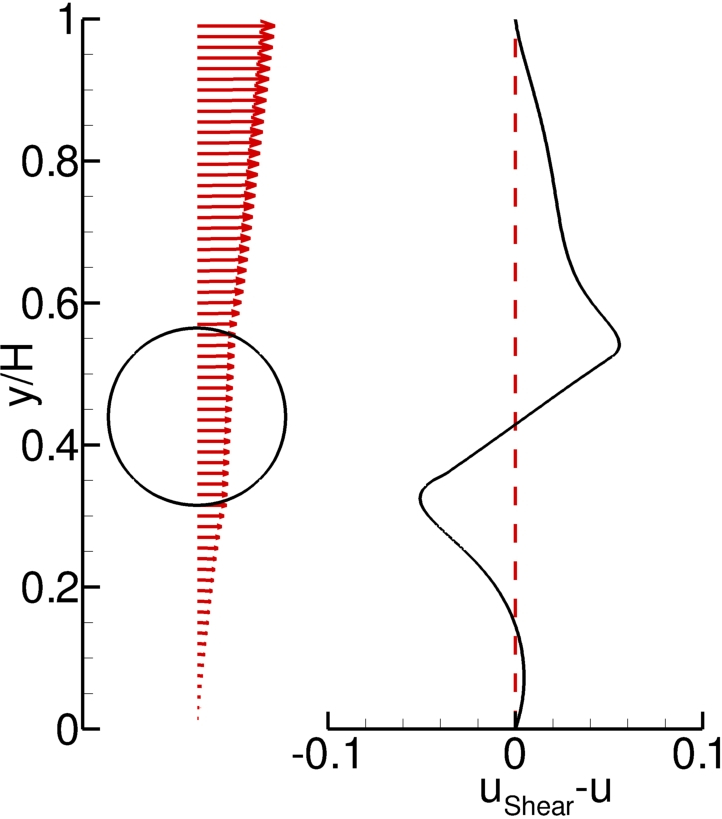}\label{profVel50}}
\subfigure[t=500]{\includegraphics[scale=0.18]{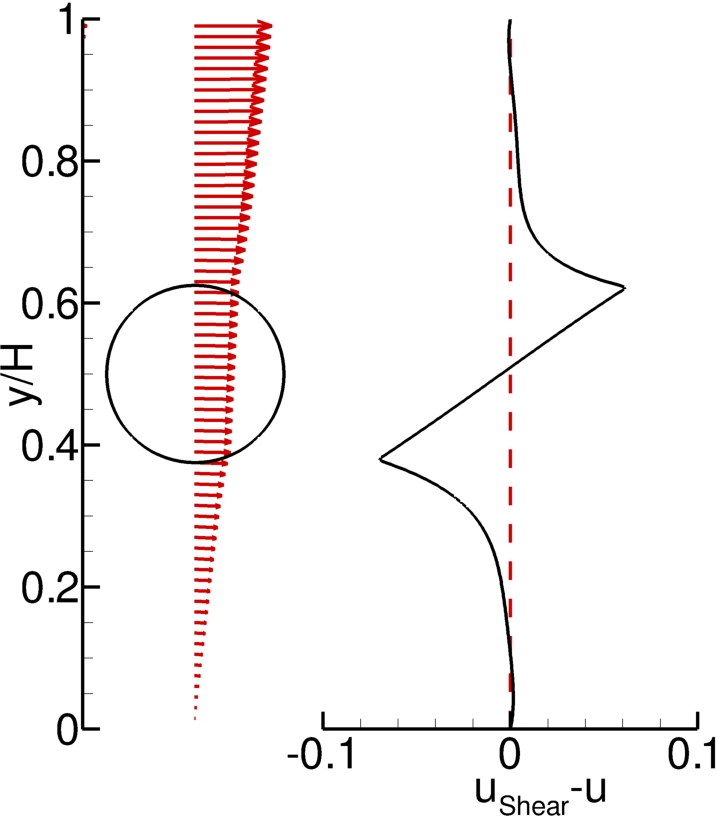}\label{profVel500}}
\caption{Local slip velocity profile taken at the axis passing by the centroid and parallel to $y$ at three different times for the particle released in $y/H=0.25$.}
\label{profVel}
\end{figure}

\begin{figure}
\centering
\includegraphics[scale=0.15]{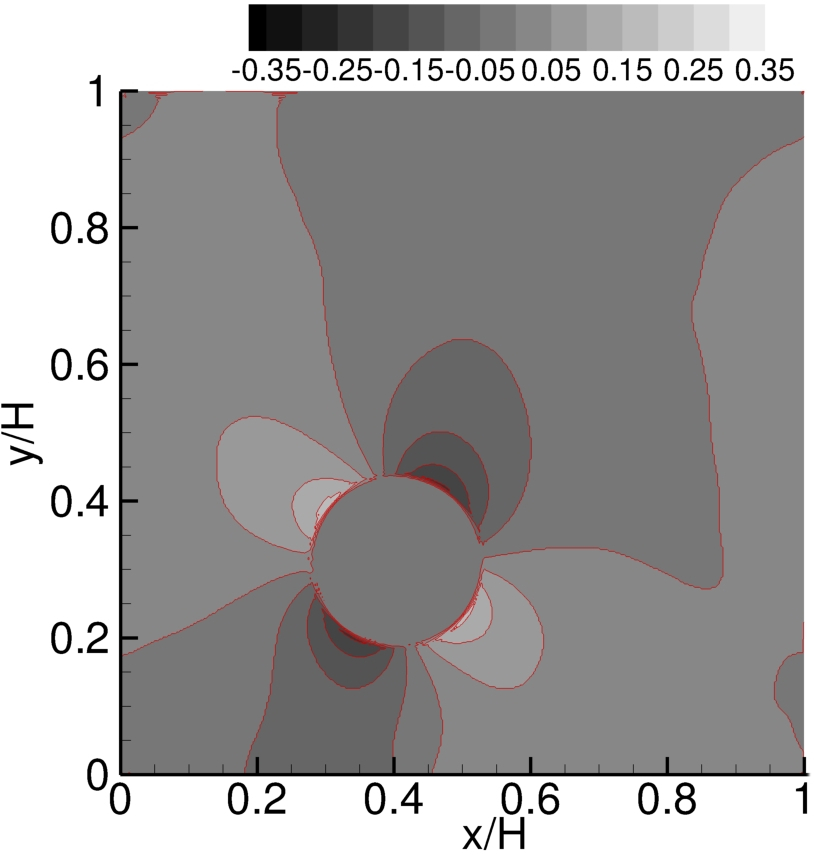}
\includegraphics[scale=0.15]{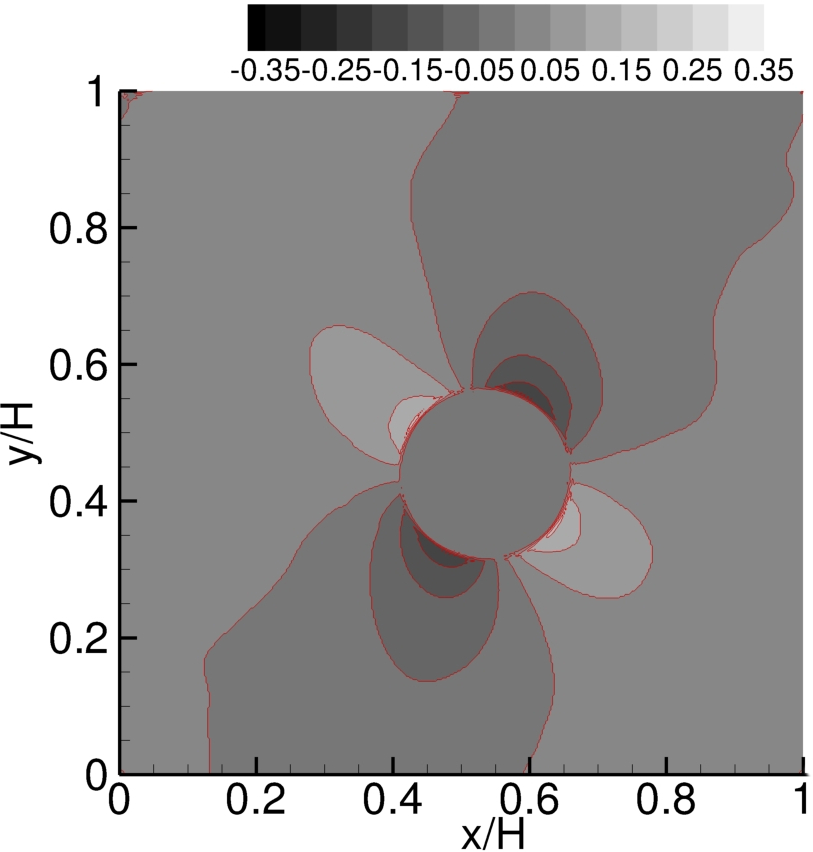}
\includegraphics[scale=0.15]{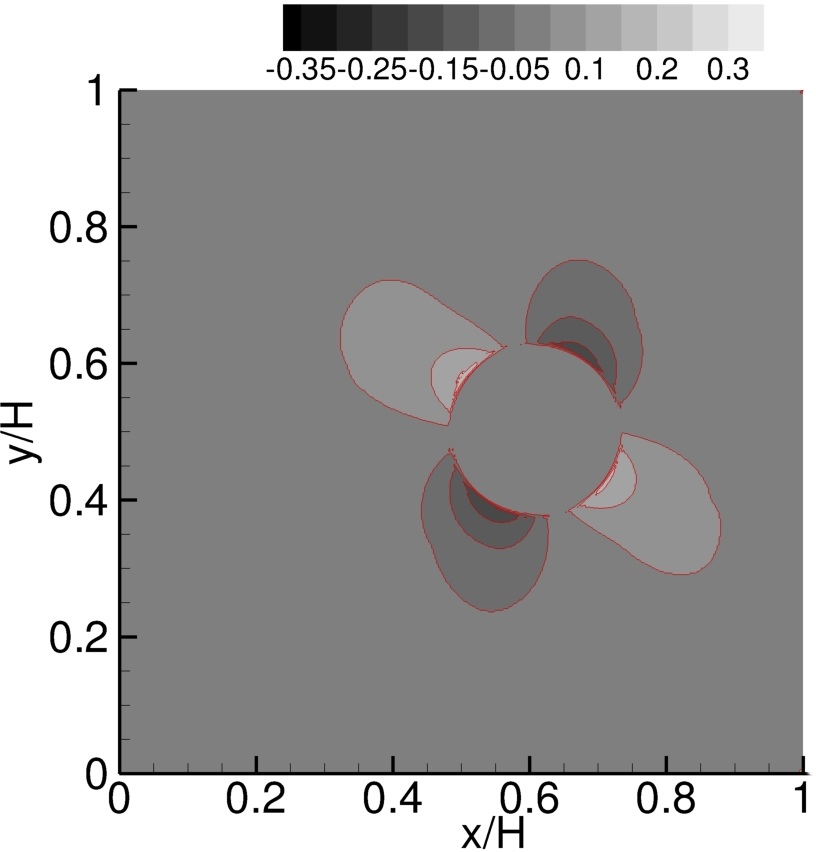}\\
\subfigure[t=10]{\includegraphics[scale=0.15]{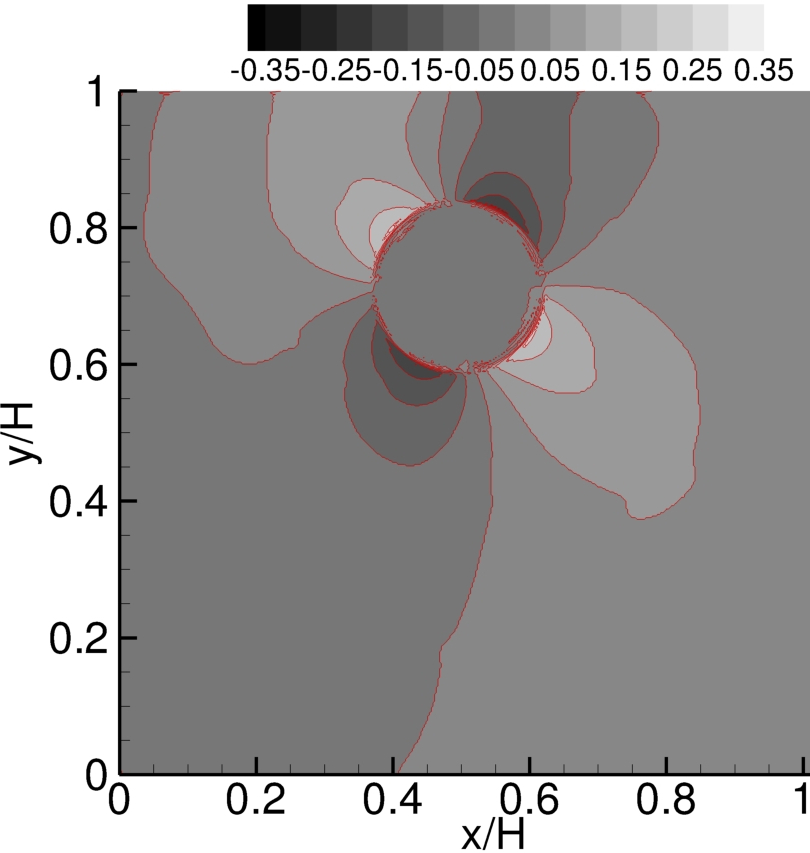}\label{10}}
\subfigure[t=50]{\includegraphics[scale=0.15]{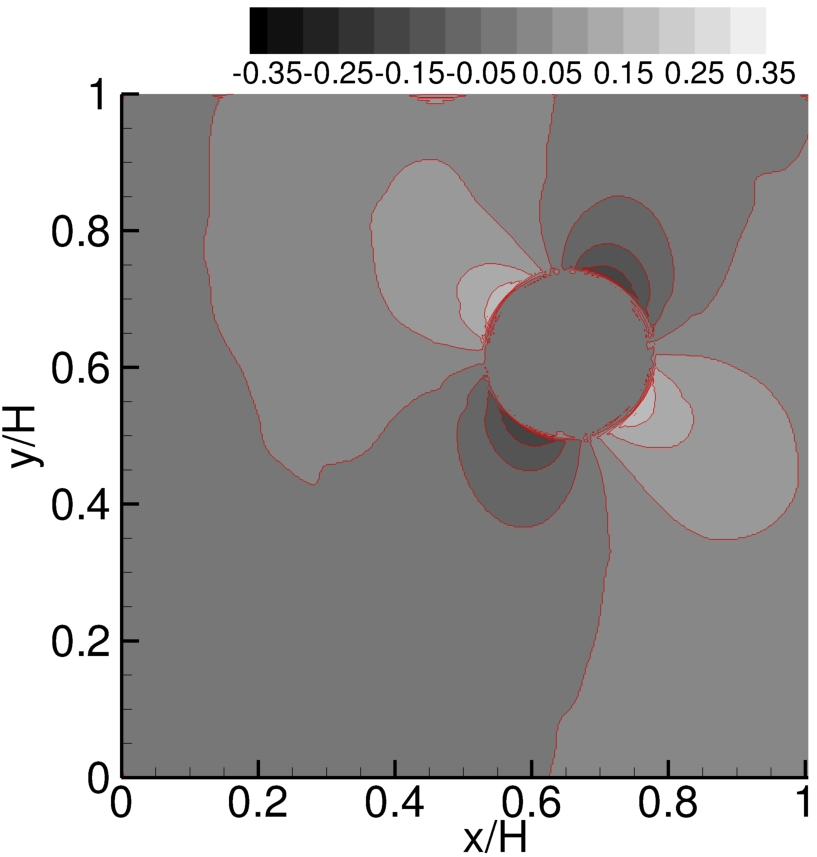}\label{50}}
\subfigure[t=500]{\includegraphics[scale=0.15]{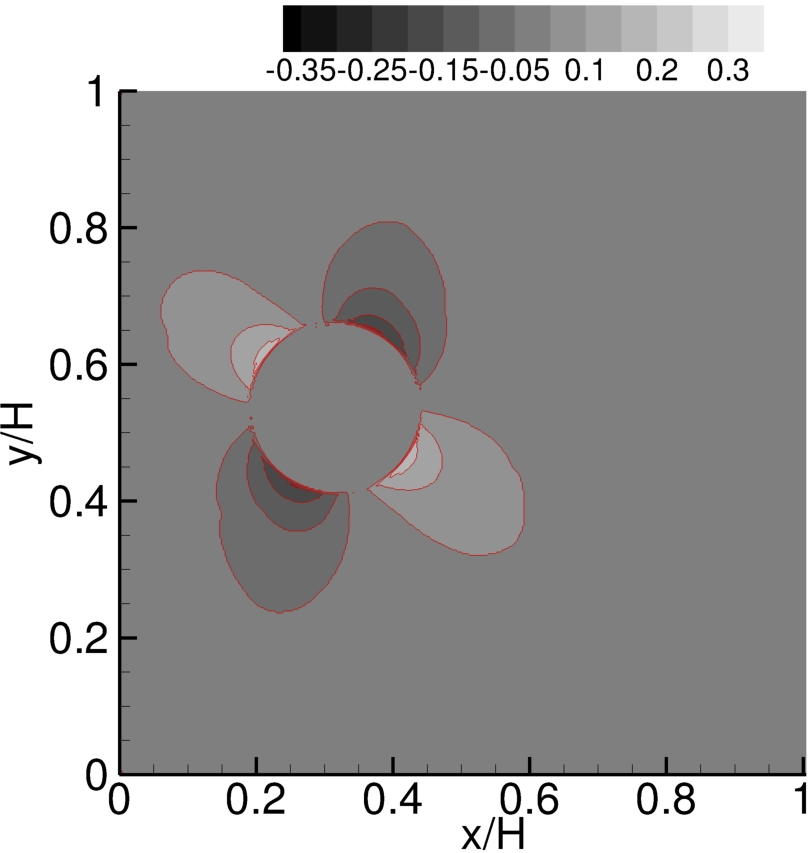}\label{500}}\\
\caption{Pressure distribution across the circular particle released in $0.25\, y/H$ (upper row) and at $0.75\, H$ (lower row) taken at three different non-dimensional time values.}
\label{pressCirc}
\end{figure}

The observed lateral drifting should be ascribed to three contributions: 
i) the inertial drift due to shear slip; 
ii) inertial drift due to rotational slip;
iii) lubrication effect due to the presence of the wall.
Through direct numerical simulation, Feng et al~\cite{feng19942} have shown that the difference between the actual particle horizontal velocity and the local undisturbed flow velocity (shear slip velocity, $u_{Slip}$) is responsible for the generation of a lateral force aiming at reducing such a difference. This force drives the particle towards flow regions where $u_{Slip}=0$. 
Specifically, a particle would move towards flow region of higher (slower) velocities if it lags (leads) the flow or, in other words, if $u_{Slip}<0\ (>0)$.
The normalised slip velocities for particles released in $y/H=0.25$ and $0.75$ are plotted in \textbf{Figure}~\ref{uSlipU}. 
For $y/H_{t=0}=0.75$, the particle leads the flow and $u_{Slip}$, larger than zero after the initial release, steadily decreases approaching zero as the particle moves towards lower flow regions. Conversely, for $y/H_{t=0}=0.25$, the particle lags the flow and $u_{Slip}$ steadily increases and approaches zero as the particle moves towards higher flow regions.
Even if an analytical form for the lift is unknown for the present case, the lift exerted by the particle appears to behave similarly to the Bretherton-Saffman lift~\cite{feng19942,saffman1965,bretherton1962}, meaning that it is proportional to Re ($u_{Slip}\,\ln(Re)^{-2}$).\\
The second contribution has been described by Joseph and Ocando~\cite{joseph2002}. 
It was shown that the angular slip velocity, $\omega_{Slip}$, defined as $\omega_{Slip}=\omega+\frac{1}{2}\dot{\gamma}$,
with $\omega$ is the angular velocity of the particle and $\dot{\gamma}$ the local shear, is responsible for a lateral force.
Notice that $-\displaystyle{\frac{1}{2}} \, \dot\gamma$ is the fluid angular velocity, so that $\omega_{Slip}=0$ at a given point
only when the particle rotates at the same velocity of the fluid. Interestingly, as the particle tends to its equilibrium position the rotational slip velocity does not approach zero, but rather settles to a value equal to about $0.03 \dot\gamma$, as shown in \textbf{Figure}~\ref{uSlipOmega}. The occurrence of a non-zero rotational slip velocity should be associated with a non-zero lateral force which would tend to dislodge the particles across the streamlines.\\
Finally, the lubrication effect generates a lateral force pushing the particle away from the wall.  
The flow through the small meniscus at the particle/wall interface is strongly reduced as compared to the flow above the particle.
\textbf{Figure}~\ref{profVel} demonstrates this mechanism showing the profiles of a convenient local slip velocity obtained at $t\times u_{Max}/L_{ref}=10,\ 50,\ 500$ along the line passing by the centroid of the particle.
This convenient local slip velocity is evaluated as difference between the undisturbed linear Couette velocity profile and the computed horizontal velocity profile, namely $u_{Shear}-u$.
It is noteworthy that, once reached the equilibrium position, the local slip velocity profile almost symmetrises (never becoming symmetrical) and so the repulsive force from the wall becomes small (never becoming null). 
This generates a net hydrodynamic pressure distribution across the particle with a higher pressure at the meniscus, which is responsible 
for a lateral force pushing the particle away from the wall.
Indeed this lubrication force reduces as the particle moves away from the wall and is balanced by the force associated with the 
residual rotational slip at the equilibrium.
\textbf{Figure}~\ref{pressCirc} depicts the hydrodynamic pressure distribution around the particles released in $y/H=0.25,\ 0.75$ 
at three time instants.
At $t=10$, a wide low pressure region appears above the particle originating a net force away from the wall. 
As time passes and the particle moves towards equilibrium, the pressure distributions along the upper and lower parts of the particle 
tend to balance leading to lower lateral forces.
Similar reasoning applies to the case of a particle released at $y/H=0.75$. 
Note that the pressure distribution in \textbf{Figure}~\ref{pressCirc} results from the contribution of all the three effects. 

\subsection{Non-circular particle transport in linear laminar flow}
\begin{figure}
\centering
\subfigure[Square particle]{\includegraphics[scale=0.15]{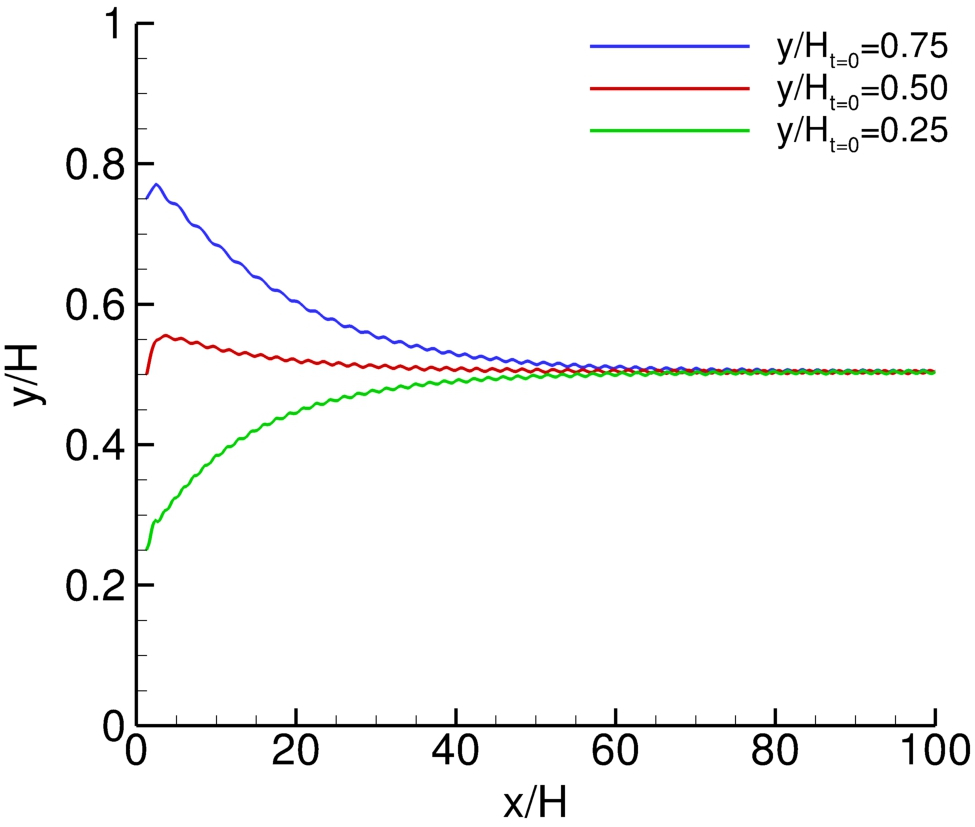}\label{SquareAll}}
\subfigure[Elliptical particle]{\includegraphics[scale=0.15]{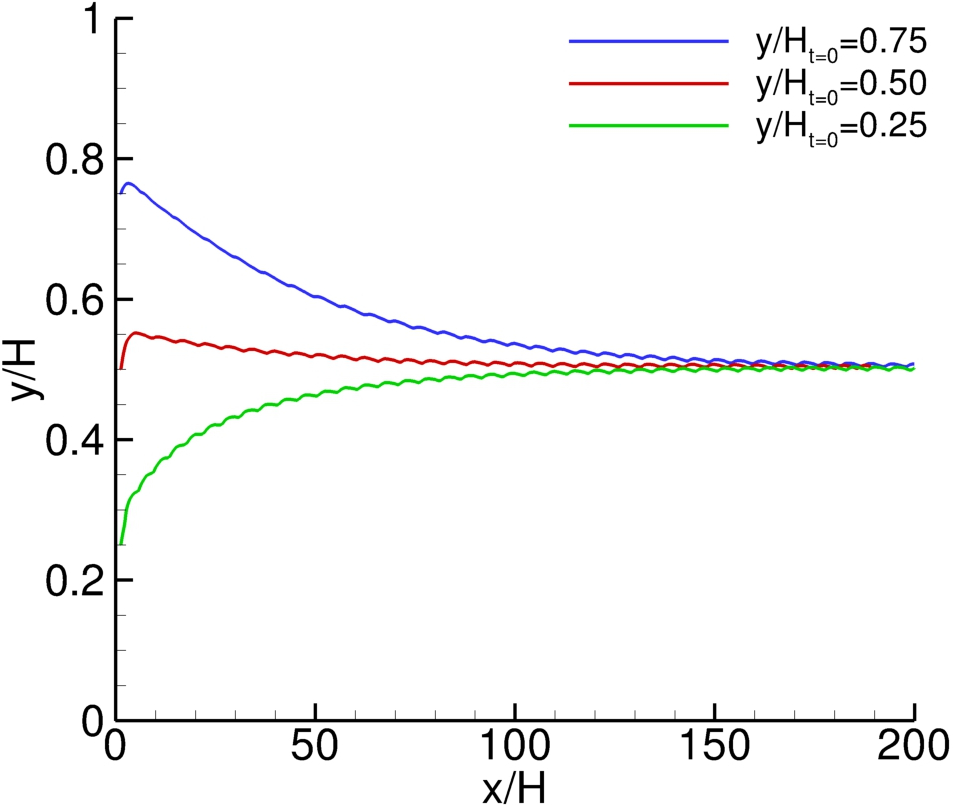}\label{EllipseAll}}
\subfigure[Triangular particle]{\includegraphics[scale=0.15]{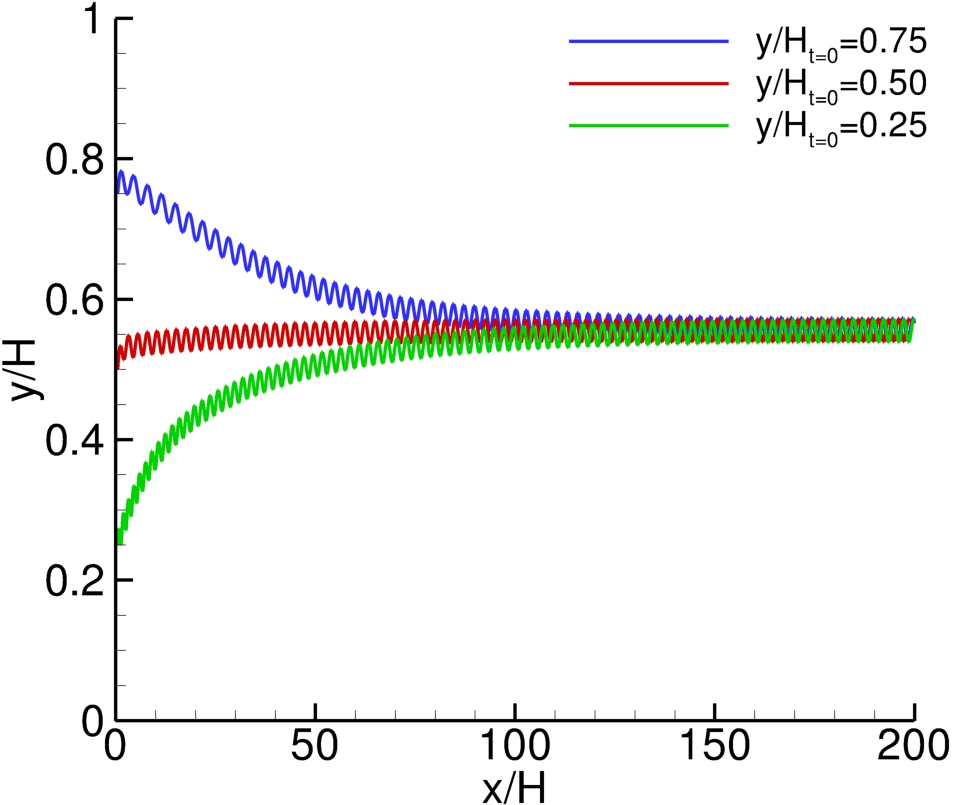}\label{TriAll}}
\caption{Particles trajectories.}
\label{TrajAll}
\end{figure}
\begin{figure}
\centering
Square particle\\
\includegraphics[scale=0.15]{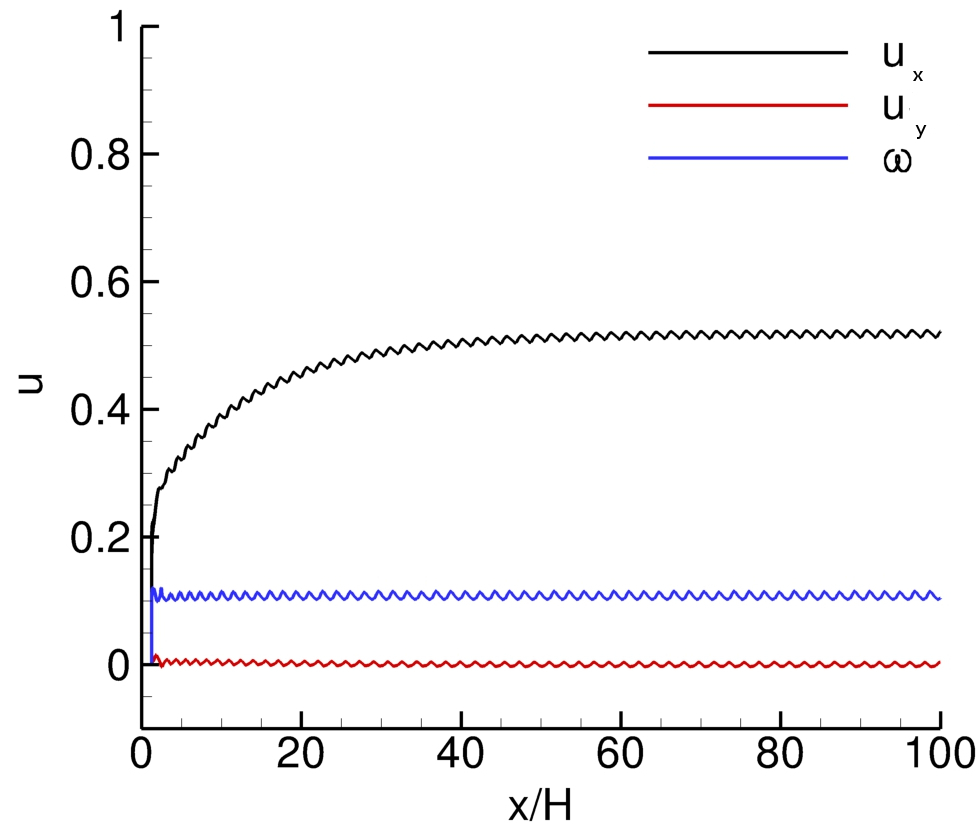}
\includegraphics[scale=0.15]{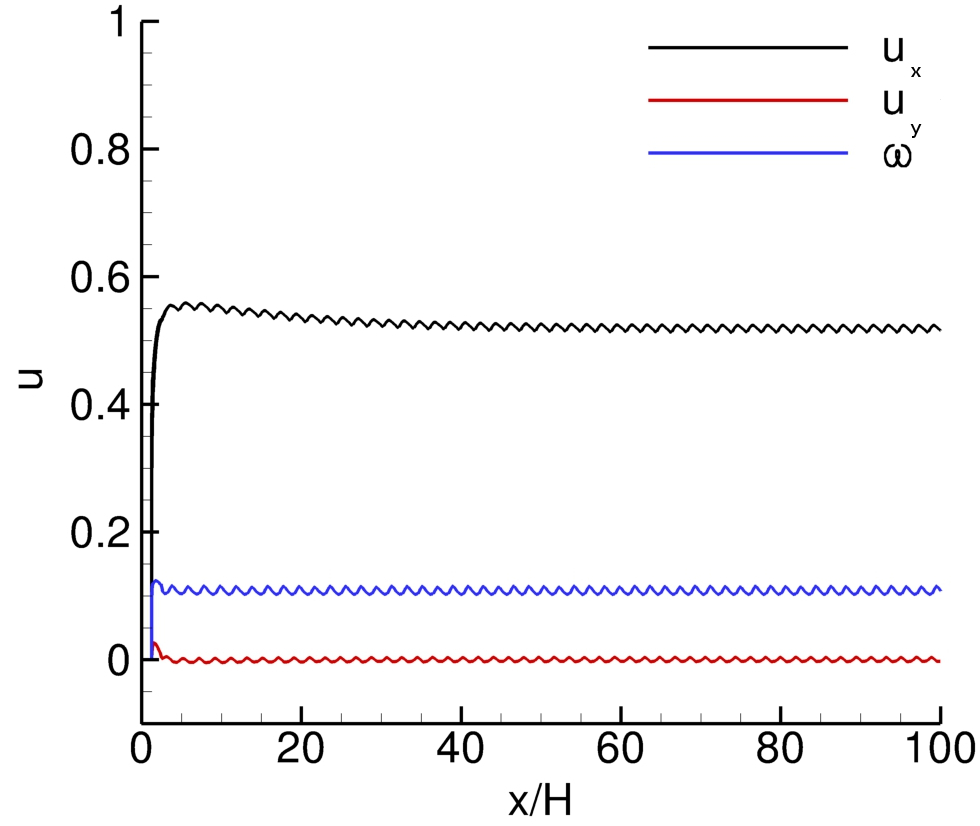}
\includegraphics[scale=0.15]{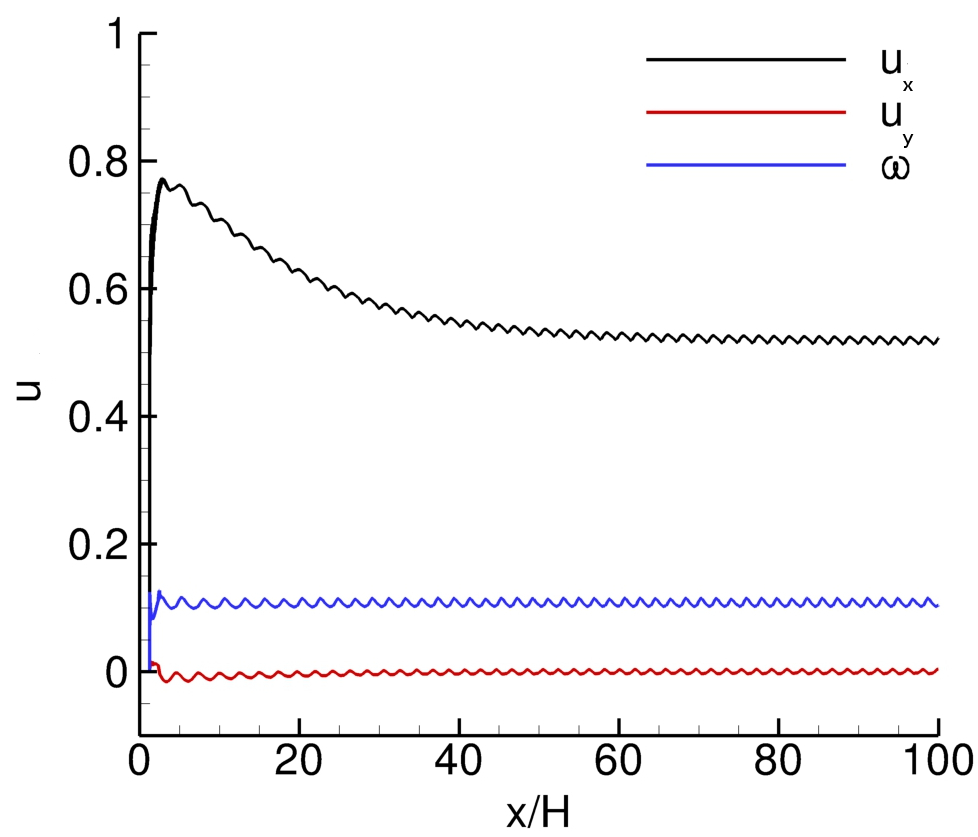}\\
Elliptical particle\\
\includegraphics[scale=0.15]{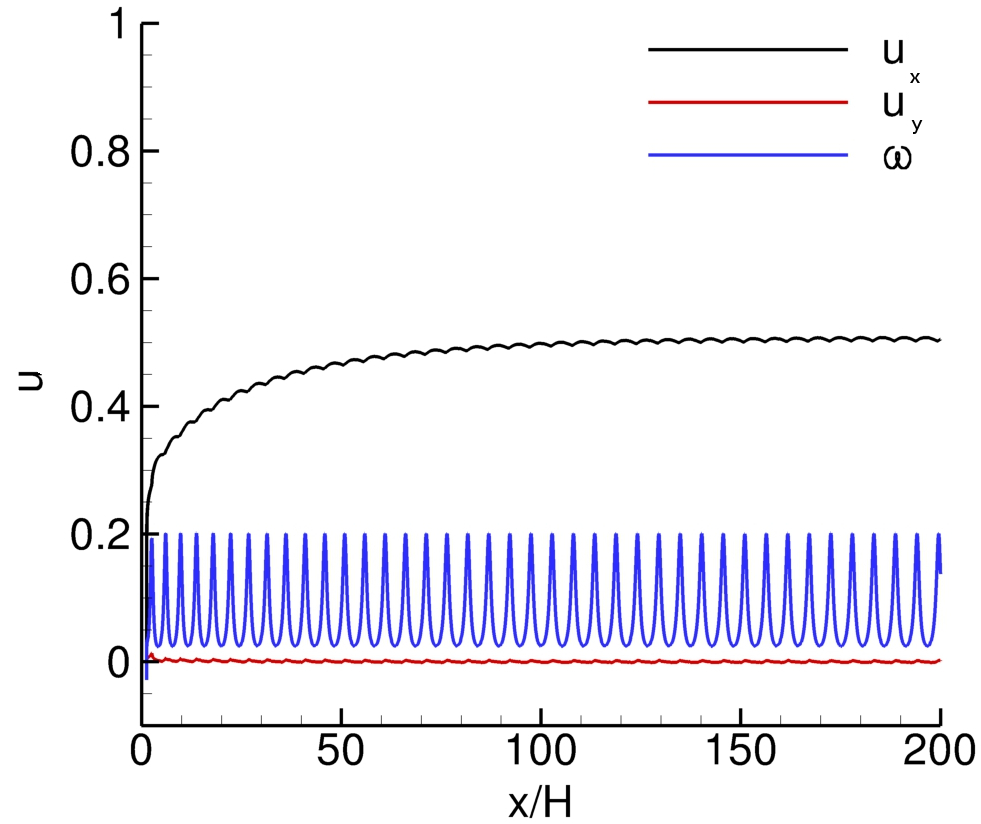}
\includegraphics[scale=0.15]{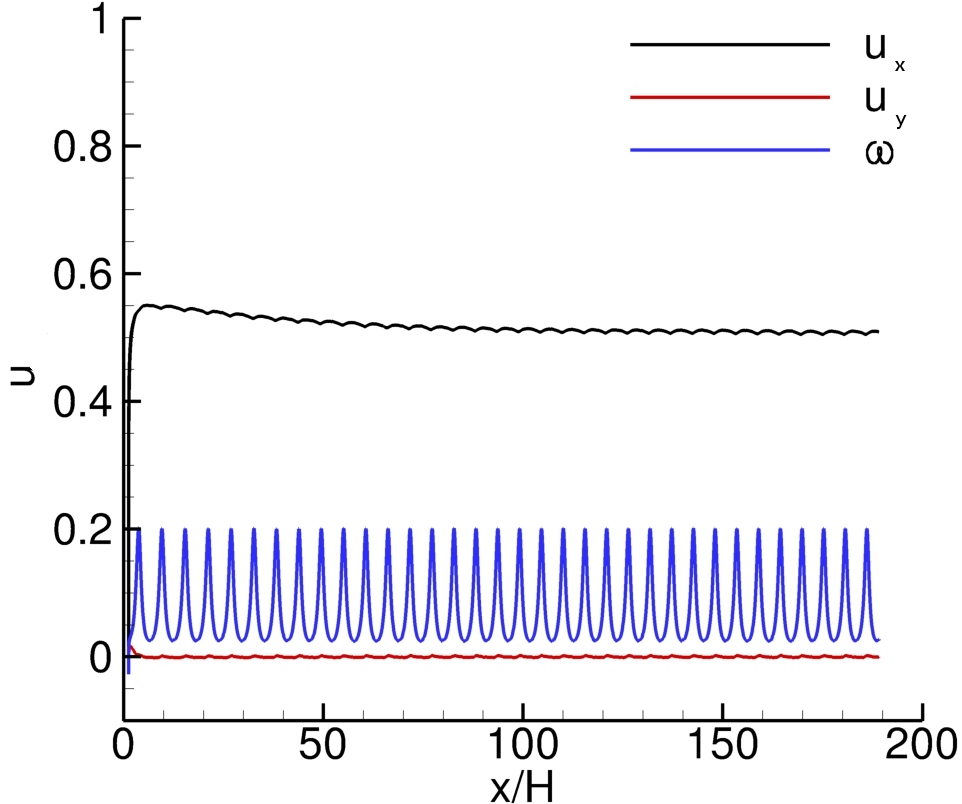}
\includegraphics[scale=0.15]{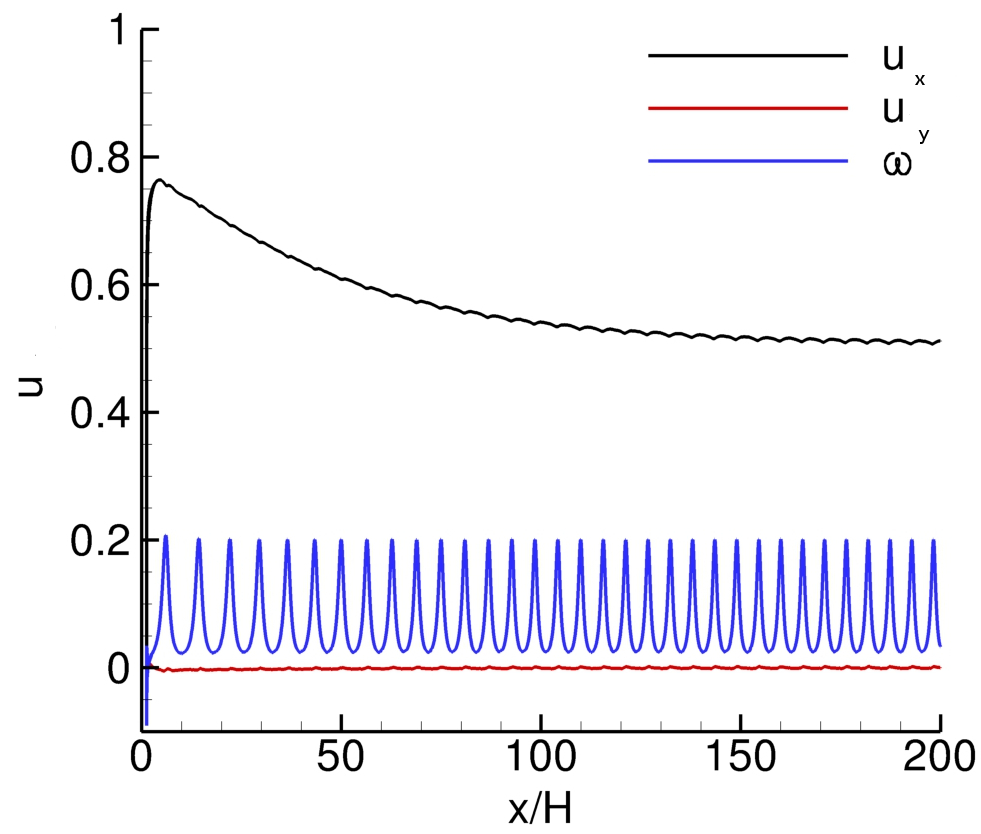}\\
Triangular particle\\
\subfigure[$y/H_{t=0}=0.25$]{\includegraphics[scale=0.15]{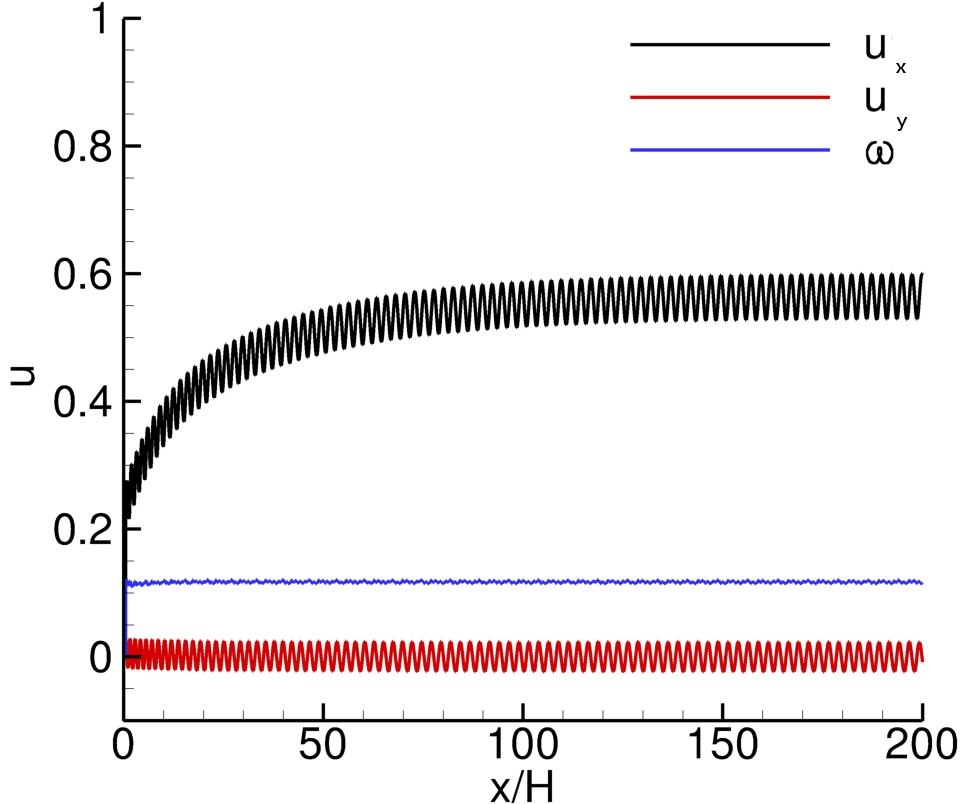}}
\subfigure[$y/H_{t=0}=0.50$]{\includegraphics[scale=0.15]{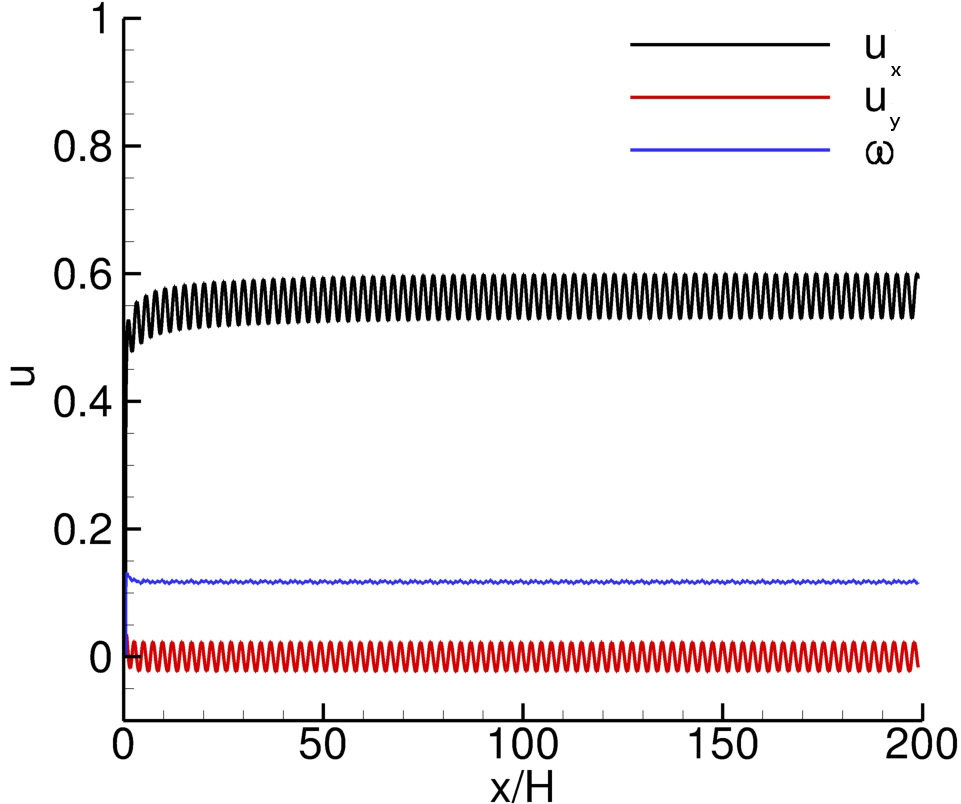}}
\subfigure[$y/H_{t=0}=0.75$]{\includegraphics[scale=0.15]{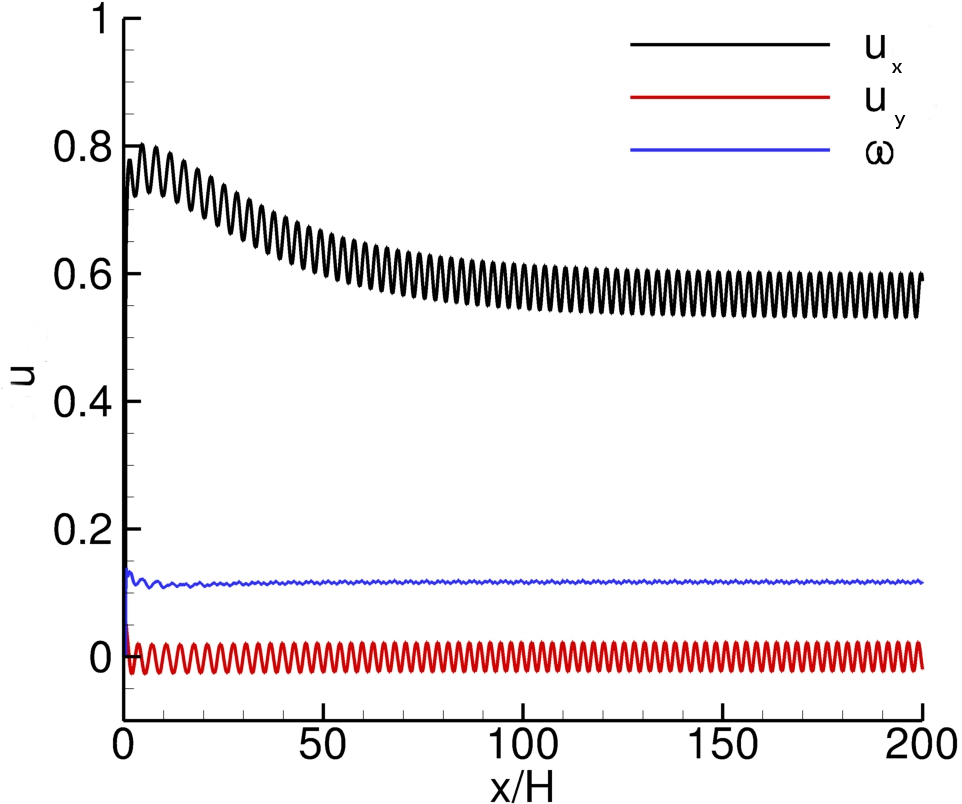}}\\
\caption{Linear and angular velocity distributions over the horizontal coordinate for the particle released at $0.25\, H$ (column (a)), 
$0.5\, H$ (column (b)), and $0.75\, H$ (column (c)).}
\label{VelAll}
\end{figure} 
\textbf{Figure}~\ref{TrajAll} depicts the trajectories of the centroids for square, elliptical and triangular particles.
As per the circular particles, the final equilibrium positions, $y_{eq}$ is independent of the initial locations. However, $y_{eq}$ is slightly affected by the particle's shape. Specifically, for square and elliptical particles $y_{eq}/H=0.50$ just as for the cylindrical particle.
Differently, the equilibrium position is $y_{eq}/H=0.53$ for the triangular particle.
A weak oscillation around $y_{eq}$ is found in the case of the square particle, see \textbf{Figure}~\ref{SquareAll}. 
This effect increases for elliptical (\textbf{Figure}~\ref{EllipseAll}) and triangular particles (\textbf{Figure}~\ref{TriAll}).
\begin{figure}
\centering
\includegraphics[scale=0.2]{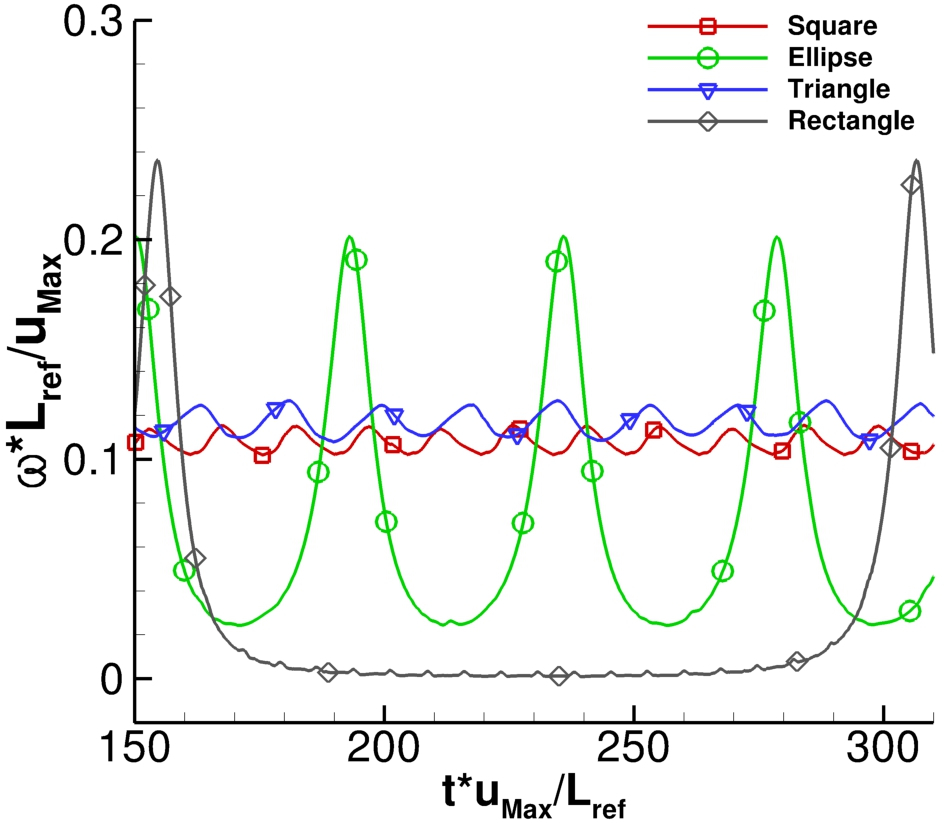}
\caption{Angular velocity distributions related to particles released at $y/H_{t=0}=0.25$.}
\label{allOmega}
\end{figure}

It is now interesting to analyse the particle oscillation around the equilibrium position.
The linear velocities, $u_x (x)$ and $u_y (x)$, shown in \textbf{Figure}~\ref{VelAll}, behave like $y(x)$ 
in \textbf{Figure}~\ref{TrajAll}, oscillating with different amplitudes and frequencies around the equilibrium position $y=y_{eq}$.
It is noteworthy that the angular velocity behaviour is determined by the geometrical characteristic of the particle. 
Indeed, $\omega$ is obtained by solving the Newton's moment equation, $I\, {\bf \dot{\omega}}=M(t)$, where $M(t)$ is the applied torque 
and $I$ the rotational inertia.
In particular, it is confirmed that particles with larger $I$ exhibit an increase in oscillation period.
\textbf{Figure}~\ref{allOmega} shows the comparison between the time variation of $\omega$ obtained considering 
square, elliptical, triangular, and rectangular particles.
The rectangular particle has a major edge length of $L_{ref}$ and an aspect ratio of $10$. The oscillation period is increased 
from $14\, t\times u_{Max}/L_{ref}$ to $147\, t\times u_{Max}/L_{ref}$ with an amplitude of about $0.018$ and $0.238$ around the 
mean value of $0.117$ when the square and the rectangular particle are respectively considered.
Moreover, one can observe the different shape of $\omega(t)$. 
The triangular and the square particles present a symmetrical oscillation around a certain mean value. On the other hand, 
the elliptical and rectangular particles present an asymmetry in both, the peak values and the peak width. Precisely, due to the elongated shape of such particles, the oscillation of the angular distribution grows with the ratio between the two main axes of the particle (\textbf{Figure}~\ref{allOmega}). 
The elongated shape leads to larger and more peaked velocities. 
In other words, the peculiar angular velocity distribution, dependent on particles shape and surface, is responsible for the oscillation around the equilibrium position in the linear velocity and consequently in the trajectories of the centre of mass of the particles.

\section*{Conclusions and future work}
\label{conclusions}

A combined Lattice Boltzmann-Immersed Boundary (LB-IB) model was proposed for predicting the transport dynamics of 
particles with different shapes, including conventional circular particles as well as less common elliptical, square, rectangular 
and triangular particles. 
A comparison between the predictions of the proposed LB-IB method and benchmark results available in the literature
for the sedimentation of circular and elliptical cylinders in a quiescent fluid, along with a mesh-refinement study for the latter
test case, showed a very good agreement, thus confirming the accuracy of the presented approach.
Similarly for the near wall dynamics of a circular cylinder in a linear laminar flow, the LB-IB method was able to predict the complex 
lateral motion of the particle in agreement with the known results of Joseph and colleagues. Specifically, it was shown that, regardless 
of the initial location and velocity, the final equilibrium position of the circular cylinder coincides with the middle stream line, with a zero slip linear velocity. 
At the considered Reynolds number (Re=20), which characterizes the vascular transport of macro-circulation, only minimal 
differences were observed between the circular and other shaped cylinders. 
After following qualitatively similar trajectories, square and elliptical cylinders found their equilibrium position on the middle stream 
line too. Differently, the triangular cylinder reached its equilibrium at 0.53~H. 
However, the angular velocities show a time-dependent behaviour with amplitude and period significantly affected by the particle shape.
Interestingly, beside the expected increase in rotational period, the more elongated particles (elliptical and rectangular) also exhibit 
a dramatic reduction in the minimum rotational velocity which, for very elongated rectangular particles, would stay at zero for almost 
the whole period. 

It should be here empathised again that the Reynolds number considered (Re=20) in the study would better represent the transport in large vessel of relatively large particles. The actual capillary transport of nanoconstructs would be characterised by significantly smaller Reynolds number (Re$<1$). Indeed, the same computational approach here proposed can be readily applied to study transport process at very low Reynolds numbers. Nevertheless, the simulations at Re=20 allowed the authors to validate accurately the model against well known test cases.

Collectively this data demonstrates that the proposed LB-IB approach can be efficiently used for predicting complex particle dynamics 
in biologically relevant flows. In the near future, three-dimensional simulations at low Reynolds number (Re$<< 1$) of rigid and deformable 
bodies would help in reproducing and predicting lateral drifting and vascular adhesion of nano/micro-particles at the micro-circulation level.
This could provide significant contributions to the field of computational nano-medicine and drug delivery.
 
\section*{Acknowledgement}\addcontentsline{toc}{section}{Acknowledgement} 
This project was supported by the European Research Council under the European Union's Seventh Framework Programme (FP7/2007-2013)/ERC Grant Agreement No. 616695.

\end{document}